\documentclass[aps,preprint,noshowpacs,nofootinbib,letterpaper,eqsecnum,preprintnumbers,superscriptaddress]{revtex4}
\usepackage{amssymb,amsmath,graphicx}
\usepackage{color}
\usepackage{subfigure}
\usepackage[colorlinks=true, urlcolor=blue, linkcolor=blue, citecolor=red, hyperindex=true, linktocpage=true]{hyperref}

\newcommand{\eps}{\epsilon}

\newcommand{\kap}{\kappa}

    \newcommand{\cH}{{\cal H}}
    
    \newcommand{\cL}{{\cal L}}
    
\newcommand{\cO}{{\cal O}}    \newcommand{\cP}{{\cal P}}
    
\newcommand{\cS}{{\cal S}}

\newcommand{\pa}{\partial}

\newcommand{\rar}{\rightarrow}

\newcommand{\be}{\begin{equation}}
\newcommand{\ee}{\end{equation}}
\newcommand{\bea}{\begin{eqnarray}}
\newcommand{\eea}{\end{eqnarray}}

\newcommand{\comment}[1]{{\bf\color{blue}[#1]}}
\newcommand{\commentr}[1]{{\bf\color{red}[#1]}}
\renewcommand{\comment}[1]{}
\renewcommand{\commentr}[1]{}

\begin{document}


\title{Pairing induced superconductivity in holography}
\author{Andrey Bagrov}
    \email{bagrov@lorentz.leidenuniv.nl}
\affiliation{Institute Lorentz, Leiden University, PO
  Box 9506, Leiden 2300 RA, The Netherlands}

\author{Balazs Meszena}
  \email{meszena@lorentz.leidenuniv.nl}
\affiliation{Institute Lorentz, Leiden University, PO
  Box 9506, Leiden 2300 RA, The Netherlands}

\author{Koenraad Schalm}
  \email{kschalm@lorentz.leidenuniv.nl}
\affiliation{Institute Lorentz, Leiden University, PO
  Box 9506, Leiden 2300 RA, The Netherlands}
\affiliation{Department of Physics, Harvard University, Cambridge MA 02138, USA}

\begin{abstract}
We study pairing induced superconductivity in large $N$ strongly coupled
systems at finite density using holography. In the weakly coupled dual gravitational
theory the mechanism is conventional BCS theory. An IR hard wall
cut-off is
included to ensure that we can controllably address the dynamics of a
single confined Fermi surface. We address in detail the interplay
between the scalar order parameter field and fermion pairing.
Adding an explicitly dynamical scalar operator with the same quantum numbers as the fermion-pair, the theory experiences a BCS/BEC crossover controlled by the relative scaling dimensions. We find the novel result that this BCS/BEC crossover exposes resonances in the canonical expectation value of the scalar operator. This occurs not only when the scaling dimension is degenerate with the Cooper pair, but also with that of higher derivative paired operators. We speculate that a proper definition of the order parameter which takes mixing with these operators into account stays finite nevertheless.

\end{abstract}

\maketitle

\section{Introduction}

The puzzles posed by strongly correlated electron systems have
been considerably illuminated in recent years by the application
of gauge-gravity duality. This ``holography'', which translates
the challenging strongly coupled dynamics to an equivalent weakly
coupled gravitational theory in one dimension higher,  has given
qualitative new insights into quantum critical transport \cite{Herzog:2007ij,Hartnoll:2007ih},
superconductivity beyond the weak coupling
Bardeen-Cooper-Schrieffer (BCS) paradigm
\cite{Gubser:2008px,Hartnoll:2008vx,Hartnoll:2008kx}, and
non-Fermi liquids \cite{Liu:2009dm,Cubrovic:2009ye}.

A simple way to pose the challenge of strongly coupled systems is that
the familiar weakly coupled particles no longer exist as controlled excitations
in this regime of the theory. Our microscopic understanding of the
observed macroscopics in condensed matter usually rests on the
notion of an electron(ic quasi)-particle --- a charged spin 1/2 fermion --- as
the fundamental degree of freedom. The theory of Fermi-liquids and the
BCS description of superconductivity are good examples of such weakly
coupled systems. Even in strongly correlated phases,
 parts of this electron quasi-particle picture survive.
The transition from such a
strongly correlated phase to a superconducting phase is still thought
to arise from fundamental electron pairing at the microscopic
level. After all, these are the only relevant charge carriers in the
system.
The open puzzle in strongly correlated electron systems such as high
Tc superconductors is the nature of the ``glue'': the interaction that
allows pairs to form.

In this article we take this suggestion that simple pairing mechanisms
should survive in strongly coupled systems to heart. While staying
ignorant on the glue,
it is a very natural step to incorporate the
BCS theory in the holographic framework.
A straightforward reason to do so is to use this very well understood standard theory
of superconductivity as a benchmark and inroad into a deeper
understanding
of holographic fermions. Although $AdS/CFT$ models of superconductivity that have been constructed up to now are quite successful in
capturing the main universal properties of real superconductors, they describe physics on
the Landau-Ginzburg level of a scalar order parameter. In doing so it
manifestly cannot reveal details of the underlying microscopic mechanisms that drive the superconducting
instability, but it also ignores the Cooper pair origin of the order
parameter. Our specific question here is whether holographic BCS can
fill in the latter gap while being agnostic on the former, and serve
as a good foothold for further research on this topic.

The most straightforward implementation of Cooper pairing in
holography is to incorporate an attractive four-fermi interaction in
the gravitational dual theory. In essence one now has a weakly coupled
BCS interaction in the dual description of the strongly coupled
theory. Pairing instabilities in this set-up were studied in
\cite{Hartman:2010fk}, and the formation of a gap in the fermion spectral
functions in a fixed Landau-Ginzburg holographic superconductor
background, charactertistic of the broken groundstate, was shown in \cite{Faulkner:2009am};
see also \cite{Chen:2009pt}.

Both these studies consider the fermions as probes. Since then our understanding of holographic fermions
has increased and we now understand that some of the peculiar
holographic effects, in particular the non-Fermi-liquid behavior,
arise from a coupling to an interacting critical IR
\cite{Sachdev:2011ze}. We shall use that improved understanding to go
beyond the probe limit and
study the full condensation of any paired state, its subsequent
groundstate and the self-consistent gap in the fluctuations around it.
One way to fully treat the fermion physics is to approximate the
fermions
in the gravitational dual in a macroscopic fluid limit
\cite{deBoer:2009wk,Hartnoll:2010gu}. In this electron star approximation it is
possible to understand the full macroscopic features of the system as
it includes gravitational backreaction. A companion article takes
this approach \cite{LiuBCS}.
The drawback of the fluid limit is that it essentially describes a
system with infinitely many Fermi surfaces --- one for each mode in
the extra radial AdS direction. This is very unusual from a condensed matter point of view.

Here we pursue an approach that allows us to concentrate on the
dynamics of a single Fermi surface. This requires us to consider the fermions
quantum-mechanically. In the straightforward holographic set-up this
``quantum electron star'' is fraught with subtle
issues due to zero-point energy renormalization and its effect on the
gravitational background \cite{Allais:2012ye,Allais:2013lha}. From the
perspective of the
field theory side this difficulty is the interaction with the large
number of
surviving IR degrees in addition to the Fermi-surface quasiparticle.
As our first goal is to simply recover the physics of regular BCS in the dual
description, the straightforward solution is to lift these extra IR
degrees of freedom, and start with a regular confined
Fermi-liquid. This can be done by the addition of a hard-wall
\cite{Sachdev:2011ze,Allais:2012ye}. This also discretizes the
infinite number of Fermi surfaces dual to each radial mode that the
AdS theory describes. We then tune the chemical potential
such that only a single Fermi surface is occupied. This has the added
virtue that the gravitational backreaction will be small, and we
are allowed to neglect it. In this straightforward set-up the bulk AdS
computation reduces to a standard Hartree BCS calculation but with
relativistic fermions in an
``effective box'' that is spatially curved. This has several technical
consequences: working in $d=3+1$ bulk dimensions, there is an
effective spin-splitting in that the up and down spin fermions have
different Fermi-momenta
\cite{Herzog:2012kx,Seo:2013nva}. Furthermore the non-trivial
wavefunctions of the fermions enter into the gap equation.
Accounting for this, we shall show that in this hard wall model conventional BCS maps cleanly between the dual
gravitational theory and the strongly interacting field theory on the boundary.

To connect this closer to previous study \cite{Faulkner:2009am}
including the standard Landau-Ginzburg holographic
superconductor, we next allow the gap-operator to become
dynamical: i.e. we introduce a kinetic term for the scalar field
in the gravitational bulk. The interpretation of this in the dual
field theory is that we have explicitly added an additional charged
scalar operator in the theory, that can independently condense. The characteristic quantum number of
this new scalar operator in the strongly coupled critical theory is
its scaling dimension. Following the well-known AdS/CFT
dictionary, this translates into the mass of dual scalar field in
the gravitational bulk. For very high mass/dimension the
field/operator decouples and we have the conventional BCS scenario
constructed earlier. For low masses, the field/operator starts to
mix with the Cooper pair operator, and we observe a BCS/BEC
crossover. Here we find a novel result. When the operator
dimension is strictly degenerate with the that of the Cooper pair, the
expectation values of each diverge. Nevertheless their sum --- equal
to the order parameter --- and the gap stay finite. In effect the
extra scalar and the Cooper pair act as a $\pi$-Josephson pair in that
the relative phase of the condensates is opposite.\footnote{Recall
  that the absolute phase of a condensate is unobservable.}

However, when the operator dimension is degenerate with that of a higher derivative
cousin of the Cooper-pair ---
higher conformal partial wave --- there is another
resonance where the naive expectation values of each diverge.
Arguably the gap should stay finite for any value of the scaling dimension.
A direct application of AdS/CFT rules does not extract the gap cleanly and indicates that a
clearer definition of the order parameter vev is needed in the AdS/CFT
dictionary. We will address this in future work. Here we conclude by shownig that one can easily construct an expression that has the right order parameter property in that
it stays finite. This postulated gap shows a clean BCS/BEC crossover.

\section{Review of Fermion spectra in the AdS dual: spin
  splitting}
\label{Plasmino}

\begin{figure}[t!]
\begin{center}

\subfigure[]{\includegraphics[width=0.49\textwidth]{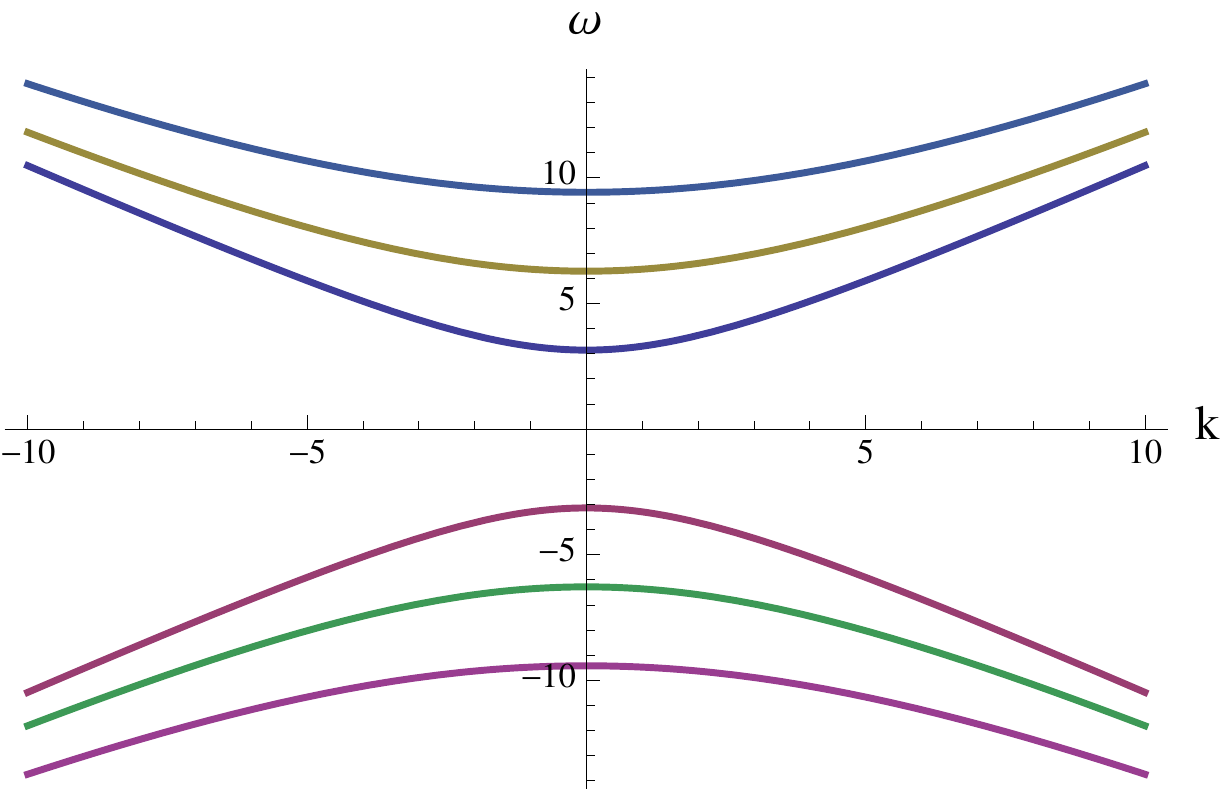}}
\subfigure[]{\includegraphics[width=0.49\textwidth]{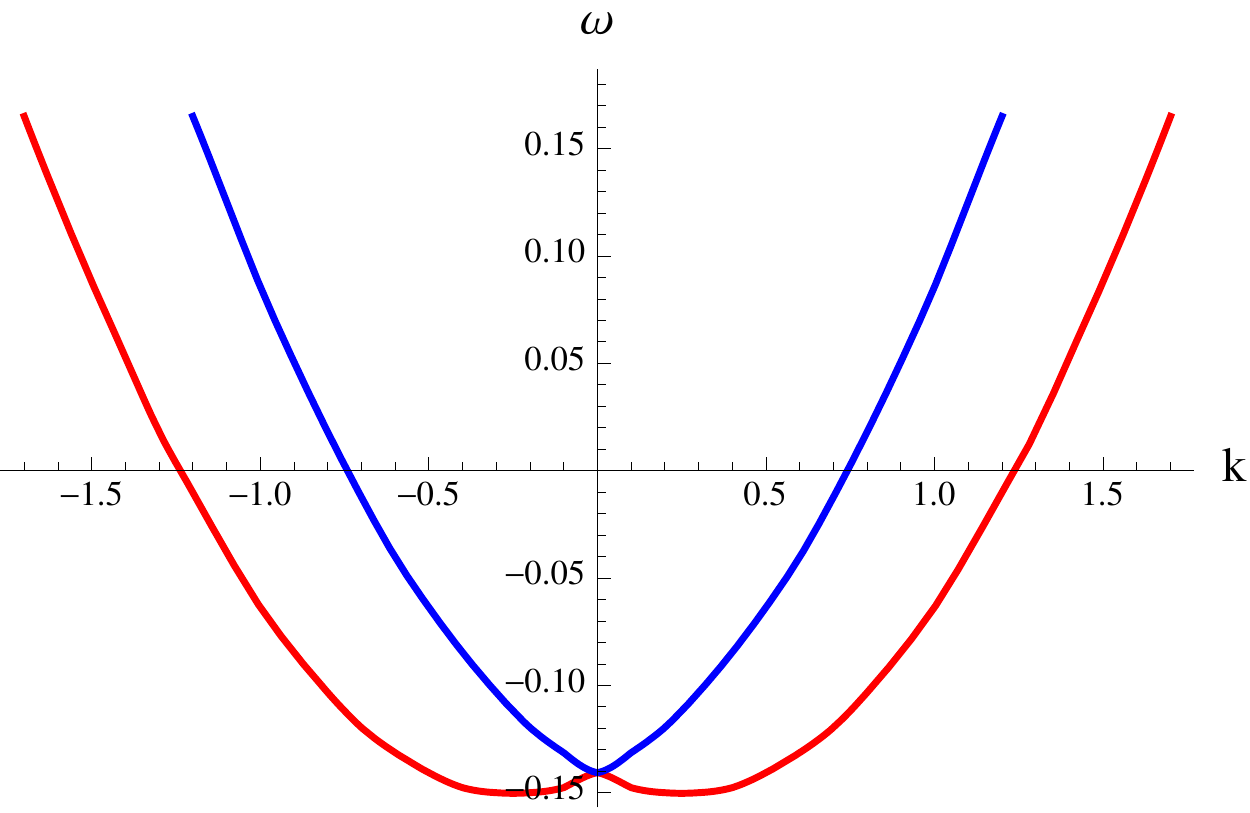}}

\caption{(a): Fermionic spectrum in the AdS-hardwall background at
zero chemical potential $z_w=1$ and $m_\Psi=1$(b): Spectrum of fermions with unit mass (and $z_w=1$) in the
 presence of externally applied electric field $q A_0(z)=4.5-2z$
 (without backreaction) \comment{UNITS}. We can observe that degeneracy
of the two spin states is resolved, and state of a minimal energy
is at non-zero momentum. The red and blue curves correspond to
positive $u_+(k)$ and negative $u_-(k)$ modes respectively. (When the electric field is self-generated by
the fermions the effect is smaller, see Fig.
\ref{fig:fermiongauge}(a))}
\label{fig:plasmino}
\end{center}

\end{figure}

To start we shall recall a lesser known point of spectra of
holographic fermions: the
spectra depend on the spin
\cite{Herzog:2012kx,Seo:2013nva}. The spectra follow from the simplest AdS
model of fermions, Einstein Dirac-Maxwell theory --- we shall add the
BCS interaction in later. The
action is
\bea
S = \frac{1}{2\kap^2}\int d^4 x \sqrt{-g}
\left(R - \frac{6}{L^2} -\frac{1}{4}F_{\mu\nu}F^{\mu\nu}+\overline{\Psi} \Gamma^\mu
D_\mu \Psi - m_\Psi \overline{\Psi} \Psi\right),
\label{eq:Dirac-action}
\eea
Here the covariant derivative equals
$D_\mu=\pa_\mu+\frac{1}{4}\omega_\mu^{ab}\Gamma_{ab}-iqA_\mu$, and
$\overline{\Psi}=i\Psi^\dagger\Gamma^0$. For the background we
choose a pure $AdS_4$ spacetime with AdS radius $L$ equal to one, and cut-off by a hard wall at a finite
value of the holographic direction $z=z_w$.
\begin{align}
  ds^2=\frac{1}{z^2}\left(-dt^2+dz^2+dx^2+dy^2\right),\,\,\,\,z\in[0,z_{w}], \label{eq:AdS-m}
\end{align}
We shall consider a large charge $q \gg \kappa$ where it is consistent to ignore gravitational
backreation. The cut off
at $z_{w}$ plays a double role. Together with the AdS potential
well, it renders the interval along the holographic coordinate
$0<z<z_w$ effectively finite. This leads to quantization of
fermionic energy bands $\omega_n(k)$ (where $n$ is the discrete
band number). Therefore, on the one hand, we have well-defined
sharp long living quasiparticles, and on the other hand the removal of
the geometry beyond $z_w$ corresponds to a gapping out of normally present low energy deconfined
degrees of freedom. This fundamental gap is also present in the
fermion spectra itself.  See Fig. \ref{fig:plasmino}(a).
In this set-up we can arrive at the dual description of a single Fermi
liquid by
tuning the chemical potential such that exactly one band is
partially occupied \cite{Sachdev:2011ze}. The charge density produced
by the occupied fermions backreacts on the gauge field
and its profile and the subsequent adjustment in the fermion spectra can be determined in a
self-consistent Hartree manner \cite{Sachdev:2011ze}.
Changing $z_w$ changes the size of the gap and the level spacing (larger values of $z_w$ correspond to smaller gap), but
does not affect the qualitative picture. Only for strictly infinite $z_w$ do we enter a new critical regime which requires a completely different analysis \cite{Allais:2012ye,Allais:2013lha}. We will keep $z_w$ finite throughout and therefore set $z_w=1$ for most of the remainder without loss of generality.
Since all our computations will only depend on the combination $q
A_0$, we also set $q=1$ in every numerical calculation from hereon.

As we shall review now,
due to the spin carried by the relativistic fermions there are
actually two Fermi liquids. Moreover, the (background or
self-generated) electric field provides a spin-orbit coupling that
renders them slightly non-degenerate in the curved background
geometry. In addition the lowest energy state is at a non-zero
momentum value; this is known as the plasmino mode
\cite{Herzog:2012kx, Seo:2013nva}. This non-degeneracy of the
different spin Fermi surfaces will be important in that it leads
to a more complex pairing of the fermions.

The spectrum of the fermions is given by normalizable solutions to the
Dirac equation.
Eliminating the spin connection by rescaling
\be \Psi=\left(-gg^{zz} \right)^{-1/4}\psi=z^{3/2}\psi, \label{eq:rescaling} \ee
Fourier transforming along the boundary directions, and making the
assumption that the only non-vanishing component of the vector
potential is $A_0$, the Dirac equation reduces to the eigenvalue problem
\begin{equation}
\left(i\Gamma^{0}\Gamma^{z}\partial_{z}+k_{i}\Gamma^{i}\Gamma^{0} - q A_0-i\frac{m_\Psi}{z}\Gamma^{0}\right)\psi=\omega \psi, \label{eq:Dirac}
\end{equation}
Hereinafter we use tangent-space gamma-matrices, and $i=1, 2$
refers to the boundary spatial indices.

Due to the impenetrability of the hard wall we choose the canonical
momenta to vanish at $z=z_w$: \bea \frac{1}{2}(1+\Gamma^{z})\psi(z_{w})=0~,~~ A'_0(z_{w})=0.   \label{eq:boundcond} \eea At the boundary $z=0$ we
demand that the fermion and scalar fields are normalizable
(i.e. vanish sufficiently fast),
and the boundary value of the gauge field sets the chemical potential
in dual field theory: $A_0(0)=\mu$.

\begin{figure}[t!]
\begin{center}

\subfigure[]{\includegraphics[width=0.49\textwidth]{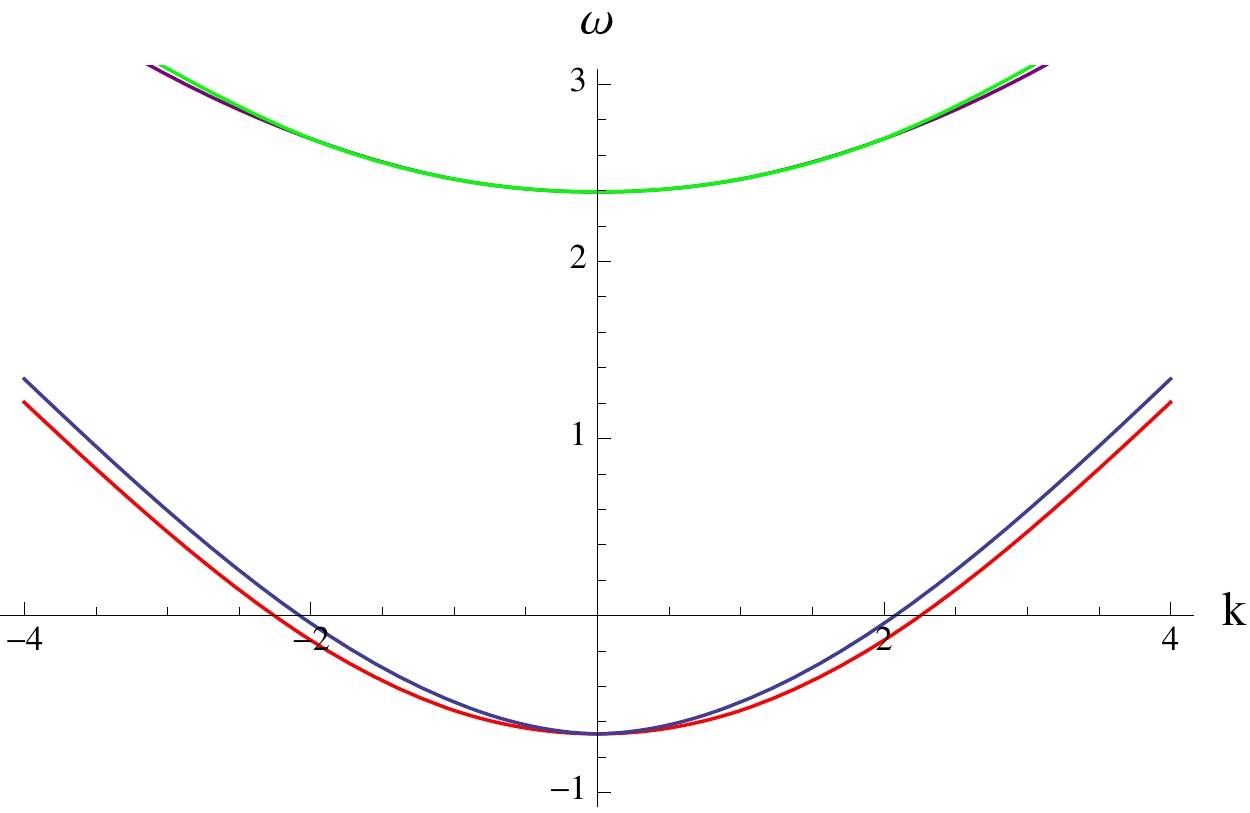}}
\subfigure[]{\includegraphics[width=0.49\textwidth]{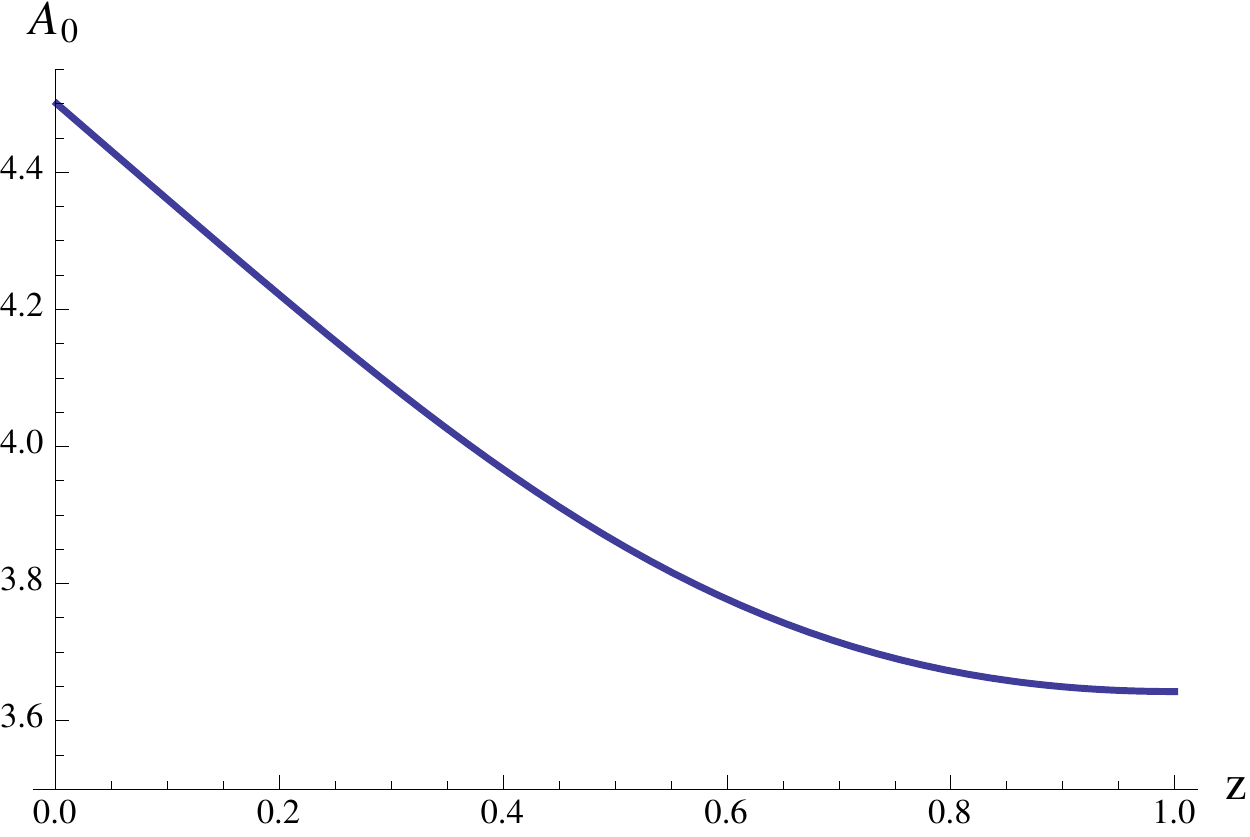}}

\caption{(a): Fermionic spectrum  in the self-consistent solution
of the fermion+gauge field system at $q \mu=4.5$, $z_w=1$ and $m_\Psi=1$. The red and blue
curves represent the modes with positive and negative eigenvalues
of $\hat{k}_{i}\Gamma^{i}\Gamma^{0}$ respectively. (b): The
profile of the gauge field sourced by the fermions.}
\label{fig:fermiongauge}
\end{center}
\end{figure}

The fermion spectra are determined together with the gauge field profile self-consistently
by (numerical)
iteration \cite{Sachdev:2011ze}:
solve the Dirac equation for a given gauge field
profile (for the initial profile $A_0(z)=\mu$). Then solve Maxwell
equations
$\nabla_{\mu}F^{\mu\nu}=-iq\langle\overline{\Psi}\Gamma^{\nu}\Psi\rangle$
with the source determined from the normalizable wave-functions. This
gives a new gauge field profile for $A_0$, etc.
the result converges to a
self-consistent solution after a few iterations (Fig.
\ref{fig:fermiongauge}).

The interesting
feature of the spectrum is that each band has a fine structure.
To understand the origin of this splitting we examine profiles of
the two spinor modes corresponding to the first band.
Fermion spectra are frequently analyzed using rotational invariance to
rotate the
momentum $k_i$ parallel to the $x$-axis and choosing an
appropriate basis of the gamma matrices one can simplify the
problem \cite{Liu:2009dm}. It will, however, be useful for us to
keep the rotational symmetry manifest.
Our objective is to separate the radial evolution of the fermion
from its spinorial structure as much as possible.
 We can solve the Dirac equation (\ref{eq:Dirac}) with the ansatz
\comment{KS ONLY: CHECK}
\begin{equation}
\psi_{\pm}(z)=A_{\pm}\left(z, \lvert \vec{k} \rvert \right)u_{\pm}\left(\hat{k}_i \right)+B_{\pm}\left(z, \lvert \vec{k} \rvert \right)\Gamma^0u_{\pm}\left(\hat{k}_i \right),
\end{equation}
where $A_{\pm}\left(z, \lvert \vec{k} \rvert \right)$ and $B_{\pm}\left(z, \lvert \vec{k} \rvert \right)$ are functions of
the radial coordinate and $u_{\pm}\left(\hat{k}_i \right)$ are
spinors (with unit norm) independent of $z$. The latter are
defined by the following properties \bea
\Gamma^{z}u_{\pm}\left(\hat{k}_i \right)=u_{\pm}\left(\hat{k}_i
\right)~,~~ \hat{k}_{i}\Gamma^{i}\Gamma^{0}u_{\pm}\left(\hat{k}_i
\right)=\pm u_{\pm} \left(\hat{k}_i \right),    \eea where
$\hat{k}_{i}$ is a unit (boundary) vector pointing to the
direction of the momentum $k_i$. In the basis (\ref{eq:basis})
(which we will use later in this paper) and with a momentum
parallel to the $x$-axis $u_{+}$ ($u_{-}$) is the spinor with only
fourth (first) nontrivial component.

The Dirac equation implies that
\begin{equation}
\left(\begin{array}{cc}
\pm \lvert \vec{k} \rvert -qA_0(z) & i\frac{m_\Psi}{z}+i\partial_{z} \\
-i\frac{m_\Psi}{z}+i\partial_{z} & \mp \lvert \vec{k} \rvert -qA_0(z)
\end{array}\right)
\left(\begin{array}{cc}
A_{\pm}\left(z, \lvert \vec{k} \rvert \right)  \\
B_{\pm}\left(z, \lvert \vec{k} \rvert \right)
\end{array}\right)
=\omega \left(\begin{array}{cc}
A_{\pm}\left(z, \lvert \vec{k} \rvert \right)  \\
B_{\pm}\left(z, \lvert \vec{k} \rvert \right)
\end{array}\right).
\end{equation}
Provided the electrostatic potential is regular near the AdS boundary
at $z=0$, the asymptotic behavior of the solution is
\begin{align}
  \label{eq:1}
  \left(\begin{array}{cc}
A_{\pm}\left(z, \lvert \vec{k} \rvert \right)  \\
B_{\pm}\left(z, \lvert \vec{k} \rvert \right)
\end{array}\right)=
a \left(\begin{array}{cc}
0  \\
1
\end{array}\right) z^{-m_\Psi}+
b \left(\begin{array}{cc}
1  \\
0
\end{array}\right) z^{m_\Psi}~.
\end{align}
Normalizable solutions are those with $a=0$. Note that the scaling dimension of the original fermion is $\Delta_{\Psi}=m_\Psi+\frac{3}{2}$ and we obtained the powers of $z$ above as a result of the rescaling (\ref{eq:rescaling}). In the IR, the boundary condition (\ref{eq:boundcond}) implies that $A_{\pm}\left(z_w, \lvert \vec{k} \rvert \right)=0$.

In the absence of an electric field (i.e. $A_0(z)$ is constant), the
positive and negative modes have the same energy.
In this case we can
actually solve our problem exactly in terms of Bessel functions
\cite{Sachdev:2011ze}
\begin{equation}
\left(\begin{array}{cc}
A_{\pm,n}\left(z,\lvert \vec{k} \rvert \right)  \\
B_{\pm,n}\left(z, \lvert \vec{k} \rvert \right)
\end{array}\right)=
N_{\pm} \sqrt{z}
\left(\begin{array}{cc}
J_{m_\Psi-\frac{1}{2}}\left(\frac{j_{n}}{z_{w}}z\right)  \\
i \frac{\pm|\vec{k}|-\sqrt{\left(j_{n}/z_{w}\right)^{2}+\vec{k}^{2}}}{j_{n}/z_{w}}J_{m_\Psi+\frac{1}{2}}\left(\frac{j_{n}}{z_{w}}z\right)
\end{array}\right), \label{eq:profile}
\end{equation}
with the dispersion relation
$\omega_n=\sqrt{\left(j_{n}/z_{w}\right)^{2}+\vec{k}^{2}}-q\mu$.
Here $j_n$ is the n-th zero of the Bessel function $J_{m_\Psi-1/2}$,
and $N_{\pm}$ is the normalization constant.

However, in the presence of an electric field in the bulk ($A'_0(z)
\neq 0$) the positive
and negative modes no longer have the same energy anymore. The reason
is that the
{\em densities} of the two modes
(\ref{eq:profile}) have
 different radial profiles. The ``effective chemical potential''
 $A_0(z)$ felt
 by each mode is therefore different, if the gauge field has a
 non-trivial $z$ dependence, and this results in a different energy
 shift for the two modes (Fig. \ref{fig:plasmino}(b)).

\section{Self interacting Fermions in AdS and a bulk BCS theory}

\subsection{Majorana interaction}

To study pairing driven superconductivity
we now
add a quartic contact
fermionic interaction in the bulk of AdS:
\begin{equation}
\cL_{contact}=\frac{\eta_5^2}{m_\phi^2}z^6\left(\overline{\psi^C}\Gamma^5\psi\right)^{\dagger}\left(\overline{\psi}\Gamma^5\psi^C\right)
,\,\,\,\,
\overline{\psi}=i\psi^\dagger\Gamma^0,\,\,\,\,\psi^C=C\Gamma^0\psi^*
\label{eq:contact}
\end{equation}
$\psi^C$ here is a charge
conjugated spinor, and the $z^6$ factor is due to the rescaling \eqref{eq:rescaling}.
One can also consider the naive relativistic generalization of the
Cooper pair $\overline{\psi^C}\psi$. However
to boil down to standard BCS in non-relativistic limit, where the
coupling occurs in s-wave channel between states time-reversed to
each other, the unique Lorentz invariant
term is actually the Majorana coupling $\overline{\psi^C}\Gamma^5\psi$
(see e.g. \cite{Bertrand:2005} for details).
We therefore focus only on this term.

\begin{figure}[t!]
\begin{center}
\includegraphics[width=0.49\textwidth]{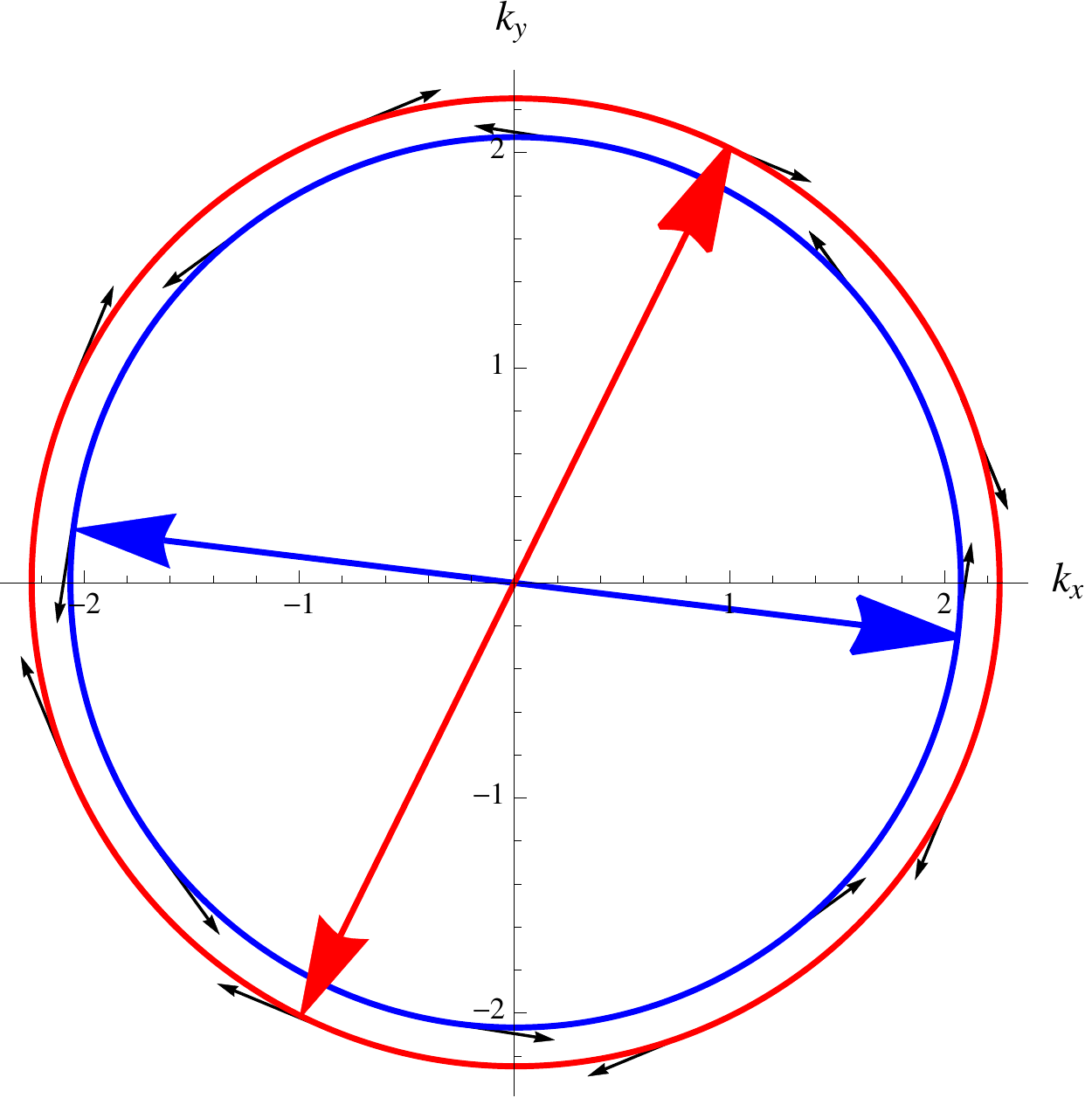}
\caption{The two Fermi surfaces and the $BCS$ pairing for the same parameters as in Fig. \ref{fig:fermiongauge}. The arrows
  indicates the direction of the spin of the modes. The pairing
  happens between opposite spins.}
\label{fig:pairing}
\end{center}

\end{figure}

As was shown in \cite{Herzog:2012kx}
the direction of the spin of each of the slightly offset modes
is perpendicular to the momenta and
the two modes have opposite spin. The zero-momentum pairing therefore occurs between
opposite spin, without any mixing of the two fermion modes, see Fig.
(\ref{fig:pairing}).

To analyze the interacting theory, we perform the standard Hubbard-Stratonovich
transformation with the introduction of an auxiliary
the scalar field $\phi(z)$ with charge $q_{\phi}=2q$ dual to the superconducting condensate.
The scalar part of the action thus takes the form
\begin{align} S = \int
d^4 x
\left( -i\eta_5^*\phi^*z^3\overline{\psi^C}\Gamma^5\psi
+\mbox{h.c.}
+ m_{\phi}^2\phi\phi^* \right)
\end{align}
This is the theory studied in \cite{Faulkner:2009am,Hartman:2010fk}
with the kinetic term for the scalar turned off. We shall reintroduce
this kinetic term in section \ref{sec:fully-inter-syst}.

\subsection{Nambu-Gorkov formalism}

The resulting system differs from standard BCS in that, as before, we
are including the backreaction of the finite density fermions on the
gauge field. Assuming translational invariance in the boundary directions, and
restrict the scalar and the gauge field to depend only on
$z$-coordinate, the holographic BCS system is formed by
 \begin{align} \label{eq:BCSeom}
 -m_{\phi}^{2}\phi(z)&=-i\eta_{5}^{*}z^{3}\langle\overline{\psi^{c}}\Gamma^{5}\psi\rangle,\nonumber\\
 z^{2}A_{0}''-2q_{\phi}^{2}A_{0}\phi^{2}&=qz^{2}\langle\psi^{+}\psi\rangle.
 \end{align}
The fermionic expectation values are assumed to only depend on $z$ as
well; they are averaged over all other directions. To compute them, it is convenient to rewrite
the action in a quadratic form in terms of the Nambu-Gorkov
spinors. We choose the following basis of gamma-matrices \be
\Gamma^{0}=\left(
\begin{matrix}  i\sigma_2 & 0 \cr 0& i\sigma_2
\end{matrix} \right),\,\Gamma^{1}=\left( \begin{matrix} \sigma_1 & 0 \cr 0 &
\sigma_1 \end{matrix} \right),\,\Gamma^{2}=\left(
\begin{matrix} 0 & \sigma_3 \cr \sigma_3 & 0 \end{matrix}
\right),\,\Gamma^{3}=\left( \begin{matrix} \sigma_3 &
0 \cr 0 & -\sigma_3 \end{matrix} \right) \label{eq:basis}, \Gamma^{5}=\left( \begin{matrix} 0 &
-i\sigma_3 \cr i\sigma_3 & 0 \end{matrix} \right).  \ee
and rewrite the fermionic part
of the action as
\begin{equation}
S_{D}+S_{M}=\int d^{4}x\sqrt{g_{zz}}\left[\overline{\psi}\Gamma^{\mu}(\partial_{\mu}-iqA_{\mu})\psi-m_\Psi\overline{\psi}\psi-i\eta_{5}^{*}\phi^{*}\overline{\psi^{c}}\Gamma^{5}\psi+\mathrm{h.c.}\right]=\int d^{4}x\overline{\chi}K\chi,
\end{equation}
where the Nambu-Gorkov spinor $\chi$ equals
\begin{equation}
\chi=\left(\begin{array}{c}
\psi_{1}\\
\psi_{2}\\
\psi_{3}^{*}\\
\psi_{4}^{*}
\end{array}\right).
\end{equation}
Taking the pure AdS metric (\ref{eq:AdS-m}) explicitly, and using
rotational invariance of the problem to set $k_y=0$, the kinetic
matrix $K$ equals
\begin{equation}
\label{eq:4}
K=\left(\begin{array}{cc}
D_{11} & 2\eta_{5}\frac{\phi}{z}\sigma_{3}\\
-2\eta_{5}^{*}\frac{\phi^{*}}{z}\sigma_{3} & D_{22}
\end{array}\right),
\end{equation}
with
\begin{equation}
D_{11}=i\sigma_{2}(\partial_{0}-igA_{0})+\sigma_{1}\partial_{x}+\sigma_{3}\partial_{z}-\frac{m_\Psi}{z},
\end{equation}
\begin{equation}
D_{22}=i\sigma_{2}(\partial_{0}+igA_{0})+\sigma_{1}\partial_{x}-\sigma_{3}\partial_{z}-\frac{m_\Psi}{z}.
\end{equation}
The fermionic expectation values can be written in terms of the
Nambu-Gorkov Green's function,
which satisfies the equation
\begin{align}
i\Gamma^0K G_{\chi_{i}\chi_{j}^{+}}\left(t,\vec{x};t',\vec{x}'\right)
&\equiv
\left(i\partial_{0}-H\right)G_{\chi_{i}\chi_{j}^{+}}\left(t,\vec{x};t',\vec{x}'\right)\nonumber\\
& =i\delta(t-t')\delta\left(\vec{x}_{\perp}-\vec{x}'_{\perp}\right)\delta(z-z'). \label{eq:greeneq}
\end{align}
Note the additional factor of $i\Gamma^0$ in our
definition.

\begin{figure}[t!]
\begin{center}
\subfigure[]{\includegraphics[width=0.49\textwidth]{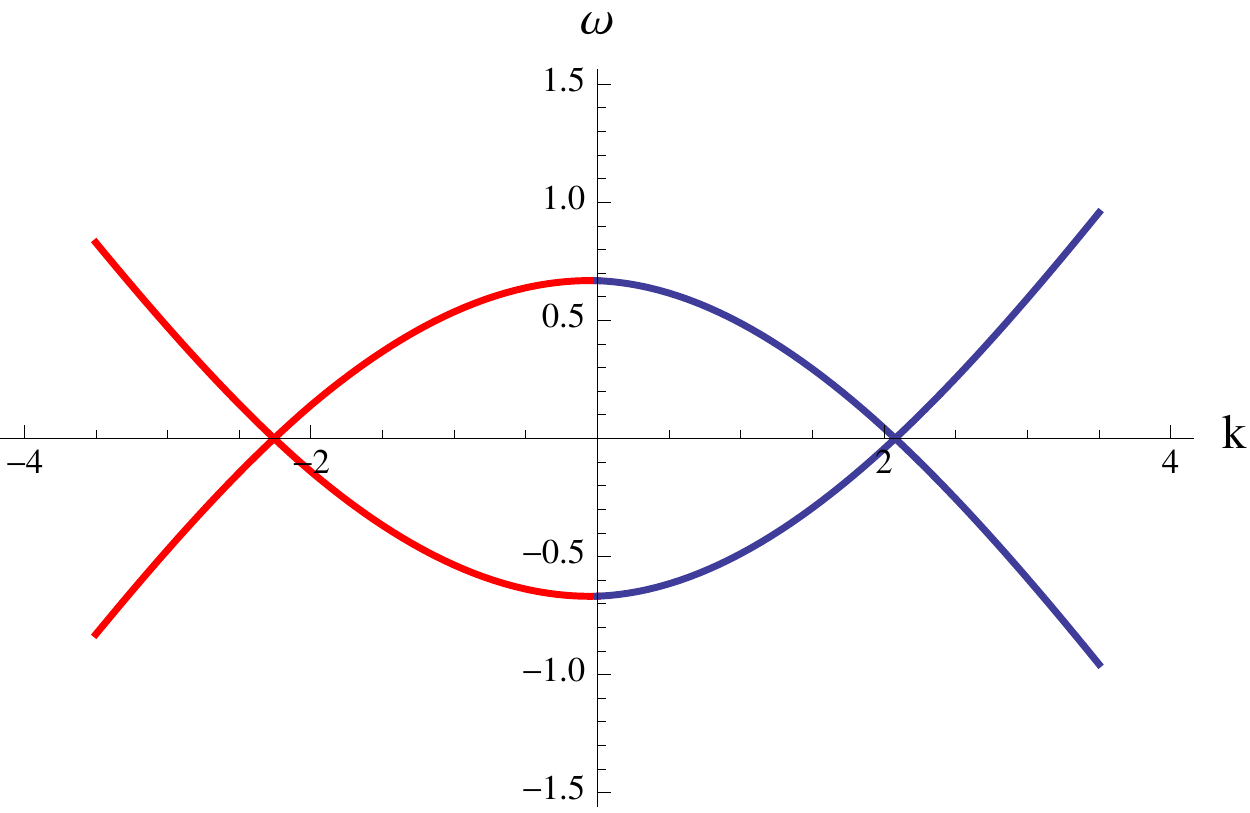}}
\subfigure[]{\includegraphics[width=0.49\textwidth]{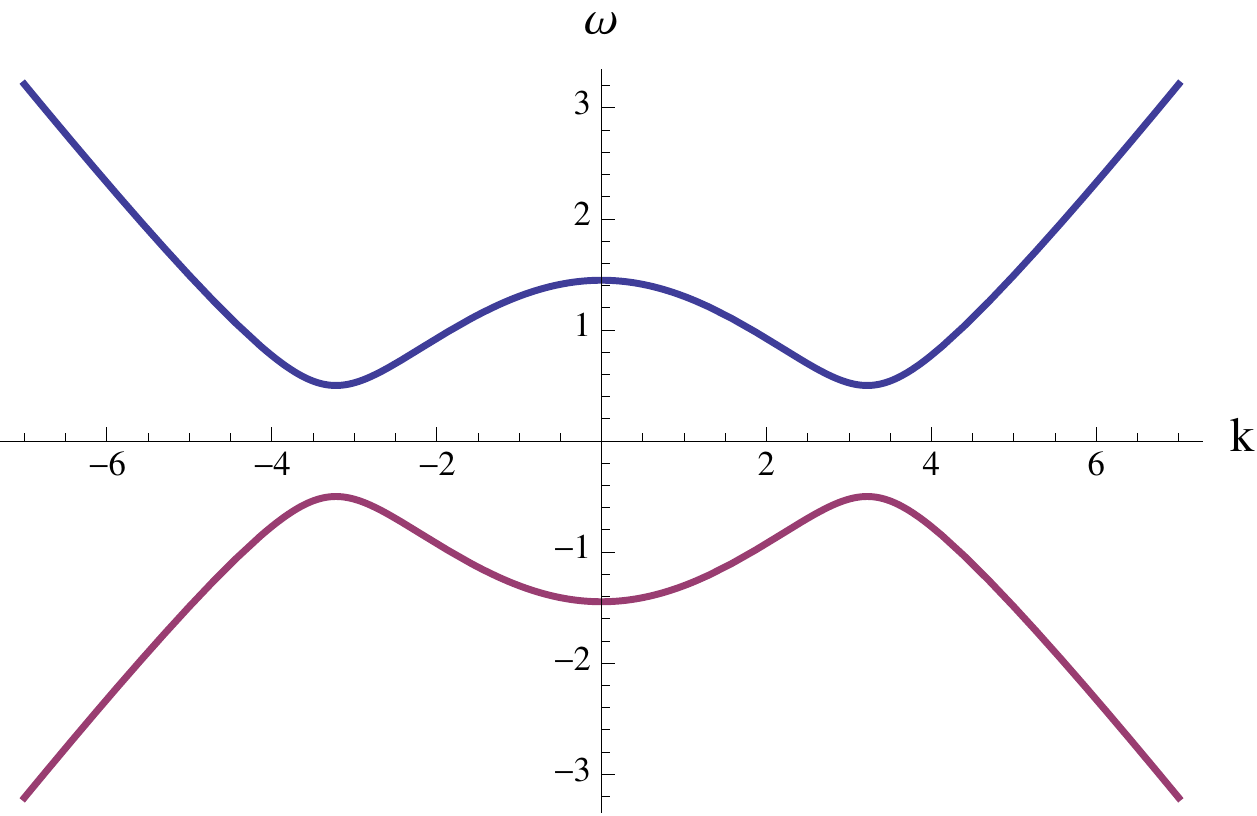}}
\caption{(a): The lower two bands from Fig. \ref{fig:fermiongauge} in the Nambu-Gorkov convention (with parameters $q \mu=4.5$, $z_w=1$, $m_\Psi=1$). (b): Energy spectrum for constant gauge field $q A_0=q \mu=4.5$ and linear fixed scalar profile $\phi(z)=z$ at $\eta_5=0.25$ ($z_w=1$, $m_\Psi=1$). The spectrum is gapped at the Fermi surface.}
\label{fig:flipped}
\end{center}
\end{figure}

We determine the Green's function by spectral decomposition. For this we solve
the Dirac eigenvalue problem in presence of both the (backreacted)
scalar and gauge field
\begin{equation}
H(i\vec{k},z)\chi_{\vec{k},n}(z)=\omega_{\vec{k},n}\chi_{\vec{k},n}(z)
\label{eq:eigenvalue}.
\end{equation}
Note, that the Nambu-Gorkov formalism flips the signs of some pieces of the
spectrum. Fig. \ref{fig:flipped}(a) shows how the two low-lying
energy bands in Fig. \ref{fig:fermiongauge}(a) look like in the
Nambu-Gorkov formalism.

It is convenient to write (\ref{eq:eigenvalue}) in terms of $(\alpha_1, \alpha_2, \alpha_3, \alpha_4)=(\chi_1, i \chi_2, \chi_3, i \chi_4)$.
In this way the redefined "Hamiltonian" H is real (but we will still denote it with $H$).

We will construct the spectrum  numerically, but it is instructive
to first consider
a toy example. We wish to show that the fermion spectrum becomes
 gapped in the presence of a condensate for $\phi$. Consider the special case when the gauge
field is constant $A_0=\mu$, and the scalar field profile is
linear $\phi(z)=z$. Then it is possible to solve the Dirac
equation exactly, and the dispersion relation (corresponding to
the first band) takes the form (Fig.
\ref{fig:flipped}(b)):
\begin{equation}
\omega^2=\left(q\mu-\sqrt{\left(j_{1}/z_{w}\right)^{2}+k^{2}}\right)^2+(2\eta_5)^2,
\end{equation}
where $j_1$ is the first zero of the Bessel-function $J_{m_\Psi-1/2}$. We
visibly see the eigenvalue repulsion responsible for the opening of a gap.

\subsection{Perturbative calculation of the scalar
  source} \label{pertcalc}

In the Nambu-Gorkov formalism it is straightforward to compute the
form of fermionic bilinears sourcing the electric and scalar
fields (see Appendix A for details).
\begin{equation}
\langle\psi^{+}\psi\rangle=\frac{1}{2\pi}\sum_{n}\int
dk|k|\left(\alpha_{k,n,1}^{2}+\alpha_{k,n,2}^{2}\right)\Theta\left(-\omega_{k,n}\right)
\label{eq:ASource}
\end{equation}
\begin{equation}
\langle\bar{\psi^{C}}\Gamma^{5}\psi\rangle=\frac{i}{2\pi}\sum_{n}\int_{-\Lambda\left(\omega_D\right)}^{\Lambda
\left(\omega_D\right)}dk|k|\left[\Theta\left(\omega_{k,n}\right)\left(\alpha_{k,n,1}\alpha_{k,n,4}-\alpha_{k,n,2}\alpha_{k,n,3}\right)\right]
\label{eq:SSourcenonpert}
\end{equation}
where the sum is over the various bands (i.e. radial modes).
The sum in the
Cooper pair condensate needs to be cut-off at a momentum scale
$\Lambda$ in order to be well-defined. This momentum cut-off corresponds to an energy cut-off $\omega_D$.\footnote{We use the conventional BCS notation for this cut-off, although there is no explicit connection to any Debye frequency here as the origin of the four-fermion interaction is left in the dark.} From now on we will be using real coupling
constant $\eta_5^*=\eta_5$.

A direct discretization of the momentum integral in
(\ref{eq:SSourcenonpert}) is not the most reliable way to numerically
computing the fermionic source for the scalar field because
contributions from different momenta are sharply peaked around the
Fermi surfaces. For higher numerical accuracy and analytical
control we solve (\ref{eq:eigenvalue}) perturbatively in the
scalar field. For this we split the Hamiltonian into an
unperturbed piece and an interaction piece $H=H_{0}+V$,
$H_{0}=H|_{\eta_{5}=0}$. The typical spectrum for the unperturbed
operator looks like the one in \mbox{Fig. \ref{fig:flipped}(a)}. With
our choice of Gamma-matrices, the eigenspinor with the unperturbed
energy $\omega_{k}^{(0)}$ and momentum parallel to the $x$-axis
takes the form {(we omit the band index)}
\begin{equation}
\alpha_{k,+}^{(0)}=\left(\begin{array}{c}
\xi_{k}\\
0
\end{array} \right)
\end{equation}
where $\xi_{k}$ is a two component spinor. There is also a mode
\begin{equation}
\alpha_{k,-}^{(0)} = \left(\begin{array}{c} 0  \\ i\sigma_2 \xi_k
\end{array}\right)
\end{equation}
with $-\omega_{k}^{(0)}$, for which only the lower two
components are non-zero. Using nearly degenerate perturbation
theory we find the matrix-element controlling the effect of the
scalar field:
\begin{equation}
V_{k}=2\eta_{5}\int_{0}^{z_{w}}dz \lvert \xi_{k}(z)\rvert^2 \frac{\phi}{z}. \label{eq:V}
\end{equation}
The new energy levels are
\begin{equation}
\omega_{\pm}=\pm\sqrt{\left(\omega_{k}^{(0)}\right)^{2}+V_{k}^{2}},
\end{equation}
so the size of the gap is $V_{k_{F}}$.
We show in the Appendix B that the scalar source has the following form in terms of the unperturbed
wave-functions {(considering only one fermion mode)}:
\begin{equation}
\label{eq:2}
\langle\bar{\psi^{C}}\Gamma^{5}\psi\rangle=-\frac{i}{4\pi}\int_{-\Lambda
\left(\omega_D\right)}^{\Lambda
\left(\omega_D\right)}dk|k|\frac{V_{k}}{\sqrt{\left(\omega_{k}^{(0)}\right)^{2}+V_{k}^{2}}}\lvert
\xi_{k}(z)\rvert^2.
\end{equation}

\subsection{Analytical study of the non-dynamical scalar: double gap equation} \label{doublegap}

Eq.\eqref{eq:2} is very similar to the standard BCS gap equation.
The
key difference is the way the spatial
profiles $\xi_k$ of the fermion wavefunctions modify both the gap $V_k$ and
the spatially varying profile of the pairing vev
$\langle\bar{\psi^{C}}\Gamma^{5}\psi\rangle$.
Since the AdS geometry together with the hard wall confine the
wavefunction,
what we have essentially done is solve a relativistic BCS in a
non-trivial potential.

There is one additional subtlety, in that the Fermi surfaces corresponding to the
up-down spin are slightly split.
Assuming, as is conventional, that the cut-off frequency is small enough, we are
allowed to approximate $V_k$ and $\xi_k$ by their values at the
Fermi surfaces. Doing so we can solve the gap equation
\begin{equation}
\phi(z)=\frac{z^{3}}{4\eta_5}\left[\gamma_1V_1\log\left(\frac{\omega_{D}+\sqrt{\omega_{D}^{2}+V_{1}^2}}{V_{1}}\right)\rho_1(z)+\gamma_2 V_2 \log\left(\frac{\omega_{D}+\sqrt{\omega_{D}^{2}+V_{2}^2}}{V_{2}}\right)\rho_2(z)\right],
\end{equation}
where $\rho_1(z)=\lvert \xi_{k_{F,1}}\rvert^2$,
$\rho_2(z)=\lvert \xi_{k_{F,2}}\rvert^2$ are the fermion wave
functions at the two distinct Fermi surfaces, and
$\gamma_{1,2}=\frac{\eta_{5}^{2}}{m_{\phi}^{2}\pi}\frac{|k_{F1,2}|}{|\omega'(k_{F1,2})|}$. A
brief inspection reveals that the gap equation only depends on the
dimensionless combinations $\frac{\eta_5}{m_{\phi}}$ and $\frac{\eta_5}{\omega_{D}}$.

In Appendix \ref{sec:simpl-gap-equat} we show that the solution of the gap equation
can be found in a form of linear combination of the two fermionic
wave functions (up to an additional $z^3$ factor)
\begin{equation}
\phi=\left(C_1\rho_1(z)+C_2\rho_2(z)\right)z^{3}.
\end{equation}
For $C_1 \gg C_2 $ ($C_2 \gg C_1 $) the condensate
profile is more similar to the wave-function at the first (second)
Fermi surface. We obtain the coefficients
\begin{equation}
C_1=(ax+b)\frac{\omega_{D}}{\eta_{5}}\exp\left(-\frac{bx+c}{\gamma_{2}}\right),
\end{equation}
\begin{equation}
C_2=\left(bx+c\right)\frac{\omega_{D}}{\eta_{5}}\exp\left(-\frac{bx+c}{\gamma_{2}}\right),
\end{equation}
where $x$ is the ratio of the two gaps $x=V_{1}/V_{2}$, satisfying the following equation
\begin{equation}
x^{2}+\left(\frac{I_{22}}{I_{12}}\frac{\gamma_{2}}{\gamma_{1}}-\frac{I_{11}}{I_{12}}\right)x-\frac{\gamma_{2}}{\gamma_{1}}=\frac{\gamma_{2}}{b}x\log x.
\end{equation}
Here $I_{11}$, $I_{22}$, $I_{12}$, $a$, $b$, $c$ are functionals of
the fermion profiles $\rho_1$, $\rho_2$, defined in (\ref{eq:I}), and (\ref{eq:a}) in Appendix B2.

\begin{figure}
\begin{center}

\subfigure[]{\includegraphics[width=0.49\textwidth]{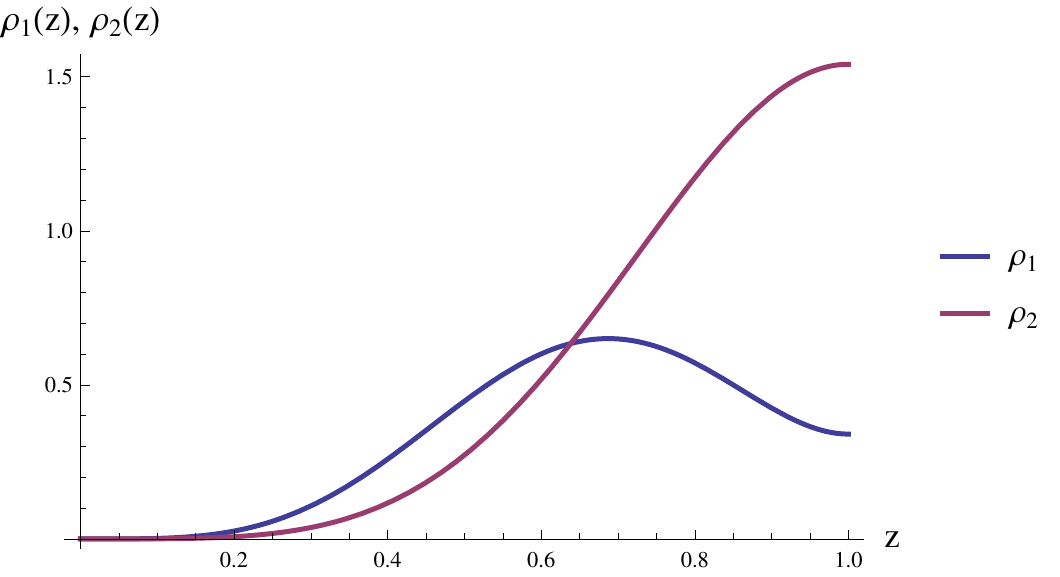}}
\subfigure[]{\includegraphics[width=0.49\textwidth]{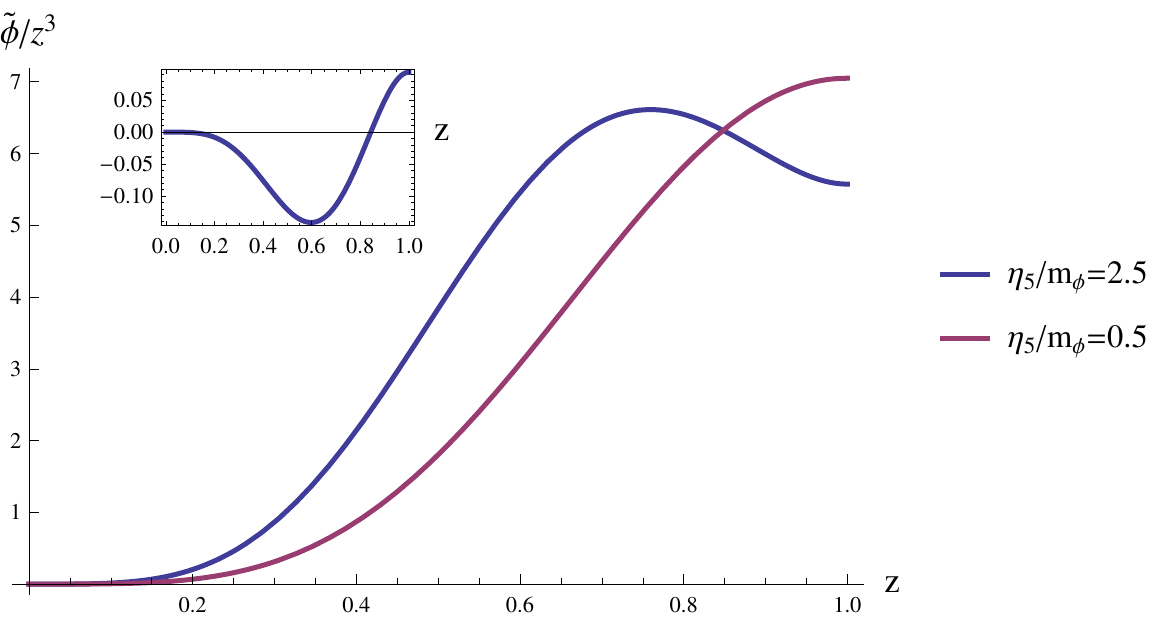}}

\caption{(a): wave function profiles of the fermions at the two Fermi surfaces ($\rho_1$, $\rho_2$) ($q\mu=4.5$, $z_w=1$, $m_{\psi}$). (b) The profiles of the stable solutions of the gap equation $\tilde \phi=\phi \exp\left(\frac{bx+c}{\gamma_{2}}\right)$ (rescaled by
$z^3$) for $\eta_{5}/{\omega_{D}}=0.5$, $\eta_{5}/m_{\phi}=0.5$ and $\eta_{5}/m_{\phi}=2.5$. Depending on the coupling the profiles are similar to the fermion wave-functions $\rho_1$, $\rho_2$. In the inset we plot the unstable solution for $\eta_{5}/m_{\phi}=2.5$ (for the other value of the coupling this mode is exponentially small).}
\label{fig:nondyn}
\end{center}
\end{figure}

\begin{figure}[t!]
\begin{center}

\subfigure[]{\includegraphics[width=0.49\textwidth]{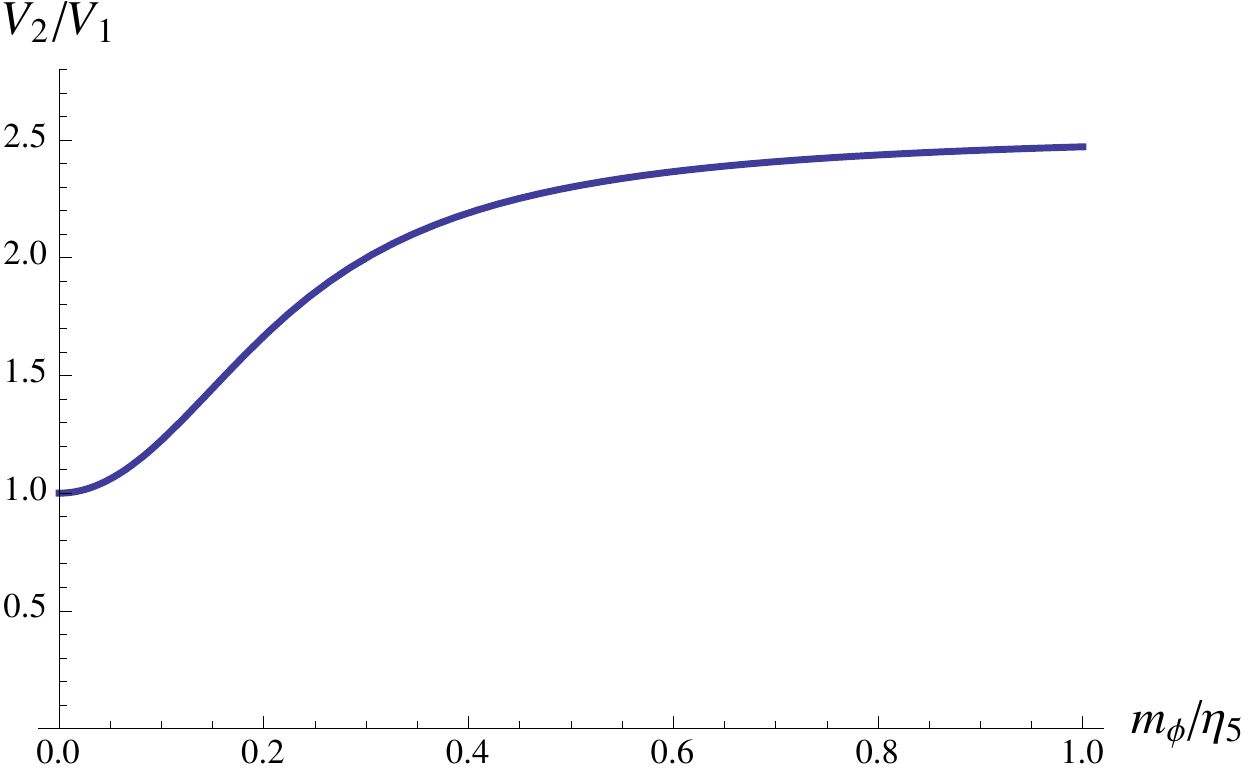}}
\subfigure[]{\includegraphics[width=0.49\textwidth]{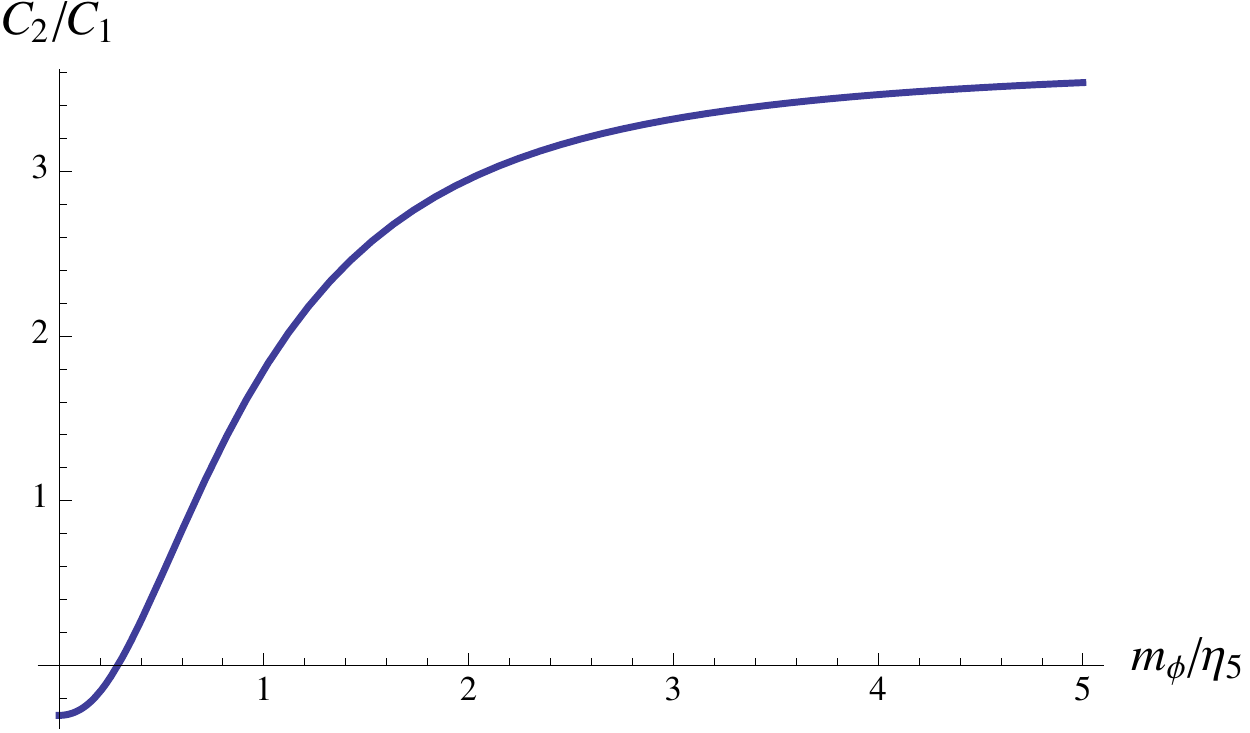}}

\caption{(a): The ratio of the gaps ($V_{2}/V_{1}$) as a function
of the inverse coupling $m_\phi/\eta_{5}$ (for fixed $\eta_5/\omega_D=0.5$).
The other parameters are as in Fig. (\ref{fig:nondyn}). For zero boson mass (or
infinite coupling) the gaps have the same size but for non-zero
mass (smaller coupling) $V_{2}$ is bigger and the ratio
converges to the value $2.56$. (b): The ratio of the coefficients
$C_2/C_1$ as a function of the inverse coupling.}

\label{fig:nondynratio}
\end{center}
\end{figure}

In Fig. \ref{fig:nondyn}(b) we show the perturbative solutions to the gap
equation for $\mu=4.5$, $q=1$, $m_\Psi=1$ and for two different
couplings. (In principle there are two solutions but one of these
contains a node and is presumably energitically unfavored).
We can see a cross-over when we tune the
coupling $\eta_{5}/m_{\phi}$ (see also Fig. \ref{fig:nondynratio}).
For small (large) coupling the profile of the condensate is
dominated by $\rho_2$ ($\rho_1$). Note that the gap at the first
Fermi-surface (with fermion wave-function $\rho_1$) is always
smaller than the gap at the second Fermi-surface.

The analysis above is all from the perspective of the bulk AdS
physics.  All the data of the dual strongly coupled field theory is
directly inferred from it. The spectral condition for a normalizable
mode is the same \cite{Sachdev:2011ze}, hence a gap in the bulk
spectra equals a gap in the boundary fermion spectrum.
The CFT order
parameter is by construction the leading non-zero component of the
fermion bilinear vev $\langle \cO_{U(1)} \rangle = \lim_{z\rar 0}
z^{-2\Delta_\Psi} \langle\overline{\Psi^{C}}\Gamma^{5}\Psi\rangle$, where
$\Delta_\Psi$ is the scaling dimension of the single trace fermionic
operator $\cO_{\Psi}$ dual to the AdS Dirac field (each normalizable
fermion wavefunction behaves as $z^{\Delta_\Psi}$)
\cite{Bolognesi:2011un,Bolognesi:2012pi}.
We thus neatly see how a bulk BCS coupling holographically encodes
standard BCS in the dual CFT.

\section{Fermionic ordering in holography}

To establish a closer connection to previous works \cite{Faulkner:2009am, Gubser:2009dt} on fermionic aspects
in holographically ordered ground states, we now introduce by hand a
kinetic term for the scalar field $\phi$. From the bulk perspective
this would correspond to a situation where the coherence length
(the inverse binding energy) of
the Cooper pair is smaller than the relevant cut-off. From the dual
boundary field theory perspective this corresponds to the introduction
of an explicit scalar operator of scaling dimension
\be
\Delta_{\phi}=\frac{3}{2}+\frac{1}{2}\sqrt{9+4m_\phi^2}.
\ee
We reserve the symbol $\Delta$ for the scaling dimensions of
operators. It is not to be confused with the value of the gap.
Again assuming translational invariance in the boundary directions,
the bosonic equations now take
 the form
 \begin{equation} \label{eq:AEOM}
z^{2}\phi''-2z\phi'+ z^{2}q_{\phi}^{2}A_{0}^{2}\phi-m_{\phi}^{2}\phi=-i\eta_{5}z^{3}\langle\overline{\psi^{c}}\Gamma^{5}\psi\rangle,
 \end{equation}
 \begin{equation}\label{eq:SEOM}
 z^{2}A_{0}''-2q_{\phi}^{2}A_{0}\phi^{2}=qz^{2}\langle\psi^{+}\psi\rangle,
 \end{equation}
where $q_{\phi}=2q$.
In addition one has the Dirac equation
\begin{align}
  \label{eq:3}
  K(\phi, A_0)\chi=0
\end{align}
through which one
defines the bulk expectation values on the right hand side. Here
$K(\phi, A_0)$
is the kinetic matrix in \eqref{eq:4},

The
distinction between the model with a dynamical and non-dynamical
scalar field is two-fold:
\begin{itemize}
\item[(1)] Although physically the order parameter in
the broken state cannot distinguish between a fermionic Cooper pair
origin and a condensed scalar, in this holographic model they
mathematically arise at different orders in the $1/N$
expansion. Recall that the coupling constant expansion in AdS/CFT maps
to the $1/N$ matrix expansion of the dual field theory, whereas each
AdS field is dual to a single trace composite operator. A Cooper pair
is thus dual to double trace operator in the dual field theory which
are always $1/N$ suppressed. This
distinction is the same distinction between classical spontaneous
symmetry breaking in a scalar field theory, and ``quantum pairing'' in
BCS.
\item[(2)] Physically, strictly put the scalar is an
additional degree of freedom (it will show up in the free
energy). If
the coherence length of the Cooper pair is smaller than the relevant
cut-off, one should indeed introduce this operator separately. In this
``strong coupling'' (equal to small coherence length) limit, the
dynamical scalar field can condense by itself. In the formulation here
this is controlled by
its mass. For high mass the field should decouple. This is dual to the
statement that in the dual field theory the corresponding operator
will have a very high scaling dimension and become extremely
irrelevant. All the IR dynamics is then controlled by the fermions and
we recover the standard BCS of the previous section. For low mass,
however, the boson dynamics will start to compete with the fermion
pairing and rapidly take over the symmetry breaking dynamics in the
IR.
\end{itemize}
Tuning the scalar mass therefore controls a crossover between pure BCS
theory and a classic BEC spontaneous symmetry breaking. Qualitatively
one can thus consider the mass/scaling dimension of the scalar
operator as a proxy for
the coherence length of the Cooper pair. When it is large, the
dynamics is pure BCS; as it becomes comparable to and  smaller than the relevant
cut-off, one should introduce the paired operator independently.

Writing out the spin components explicitly the full system
of equations that we are attempting to solve is
\bea
&&z^2\phi''-2z\phi'+4q^2
z^2A_0^2\phi-m_\phi^2\phi=
\frac{\eta_{5}z^{3}}{2\pi}\sum_{n}\int_{-\Lambda\left(\omega_D\right)}^{\Lambda
\left(\omega_D\right)} \!\!\!\!\!\!dk|k|
\Theta\left(\omega_{k,n}\right)\left(\alpha_{k,n,1}\alpha_{k,n,4}-\alpha_{k,n,2}\alpha_{k,n,3}\right),\nonumber \label{MainSystem1}\\
&&\hspace*{1.4in}z^2A_0''-8q^2A_0=\frac{q z^2}{2\pi}\sum_{n}\int
dk|k|\left(\alpha_{k,n,1}^{2}+\alpha_{k,n,2}^{2}\right)\Theta\left(-\omega_{k,n}\right),
\nonumber\label{MainSystem2}\\
&&\left( \begin{matrix}\partial_z-\frac{m_\Psi}{z} & -(\omega - k)-q A_0 & 2\eta_5 \frac{\phi}{z} & 0 \\
(\omega + k)+q A_0 & \partial_z+\frac{m_\Psi}{z} & 0 & 2\eta_5 \frac{\phi}{z} \\
2\eta_5 \frac{\phi}{z} & 0 & \partial_z+\frac{m_\Psi}{z} & (\omega - k)-q A_0\\
0 & 2\eta_5 \frac{\phi}{z} & -(\omega + k)+q A_0 & \partial_z-\frac{m_\Psi}{z}
 \end{matrix}\right)\left( \begin{matrix}\alpha_1 \\
 \alpha_2 \\ \alpha_3 \\ \alpha_4 \end{matrix}\right)=0 \label{MainSystem3}
\eea
Here all fields depend on only on the radial direction $z$.
For completeness we recall boundary conditions for each of the
fields. At the impenetrable hard wall all canonical momenta should
vanish. For the bosons this means
\begin{align}
  \label{eq:5}
   \phi'(z_{w})=0~,~~A_0'(z_{w})=0~;
\end{align}
for the fermions this can be achieved by the choice
\begin{align}
  \label{eq:6}
  \alpha_1(z_{w})=\alpha_4(z_{w})=0
\end{align}
At the AdS boundary, all field should be normalizable: they should
vanish as a positive power of $z$. (For two of the fermion components
this is automatic, see eq. \eqref{eq:1}).

We will approach the fully interacting scalar-fermion
system in three steps: we first set all fermions to vanish and
construct the purely scalar holographic
superconductor. Next we include fermions, but hold the BCS coupling
$\eta_5=0$; this exhibits bose-fermi competition in the
system. Finally we will analyze fully
interacting system at $\eta_5\neq 0$.
Details of the numerical calculations are discussed in Appendix C.

\subsection{Purely scalar holographic superconductor}

First, as the scalar field in our system is a fully dynamical degree
of freedom, it should condense for small enough mass even in absence
of fermions \cite{Hartnoll:2008vx,Nishioka:2009zj,Horowitz:2010jq}. This hardwall superconductor will be useful for later comparison.

Since we consider a pure hardwall
$AdS_4$ spacetime without a black hole horizon, we study a $T=0$
groundstate as a function of the mass/conformal dimension of the
scalar field/dual scalar operator. Any phase transition is therefore
of quantum origin. Note that the hard gap due to the hardwall directly
implies that the physics is the same for any temperature $T <
1/z_w$. Only when $T>1/z_w$ will the the black hole horizon become
relevant to the geometry, see e.g. \cite{CaronHuot:2006te}.

\begin{figure}[t!]
\begin{center}
\includegraphics[scale=0.6]{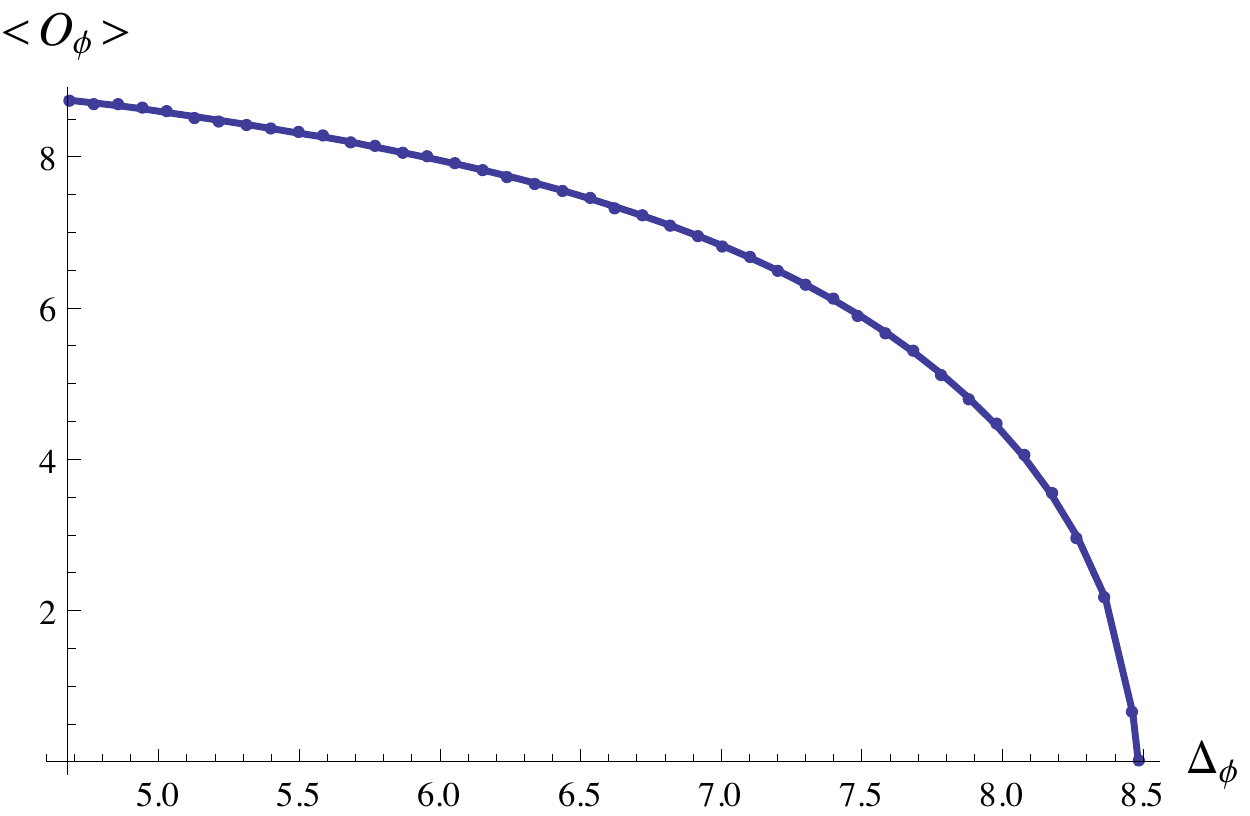}
\caption{Condensate of a scalar order parameter in the boundary theory
  as a function of scalar conformal dimension at $\mu q=4.5$, $z_w=1$, $q_\phi=2$.}
\label{fig:PureHS}
\end{center}
\end{figure}

The numerics of the pure scalar system is particularly simple as there
is no need to solve the integro-differential equations iteratively.
Varying the scalar conformal dimension we indeed find a condensate
value below a critical value (Fig. \ref{fig:PureHS}). We see a sharp
second order phase transition as expected for spontaneous symmetry breaking.
Scalar operators with smaller conformal dimensions (dual to lighter
bulk scalar fields) are more likely to condense and yield an order parameter
with higher density.

\subsection{Bose-Fermi competition}

\begin{figure}[t!]
\begin{center}

\subfigure[]{\includegraphics[width=0.49\textwidth]{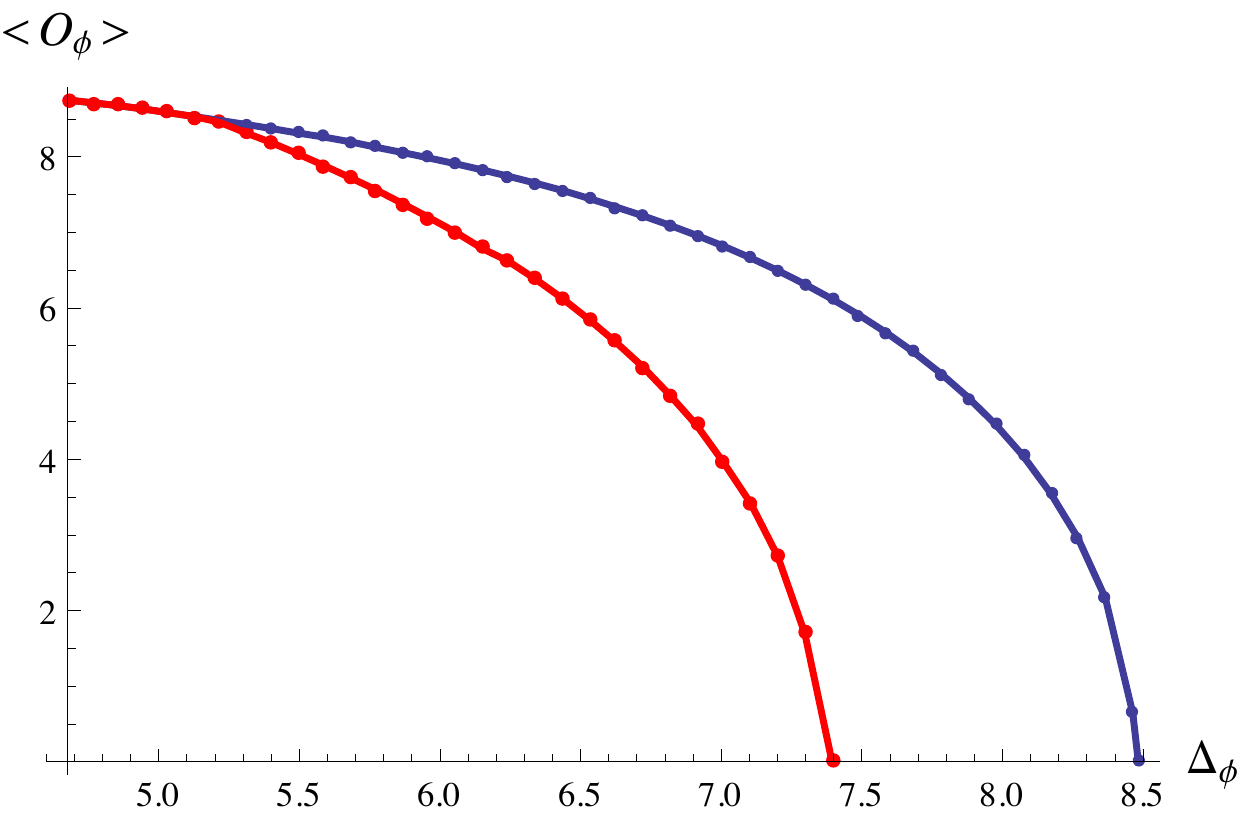}}
\subfigure[]{\includegraphics[width=0.49\textwidth]{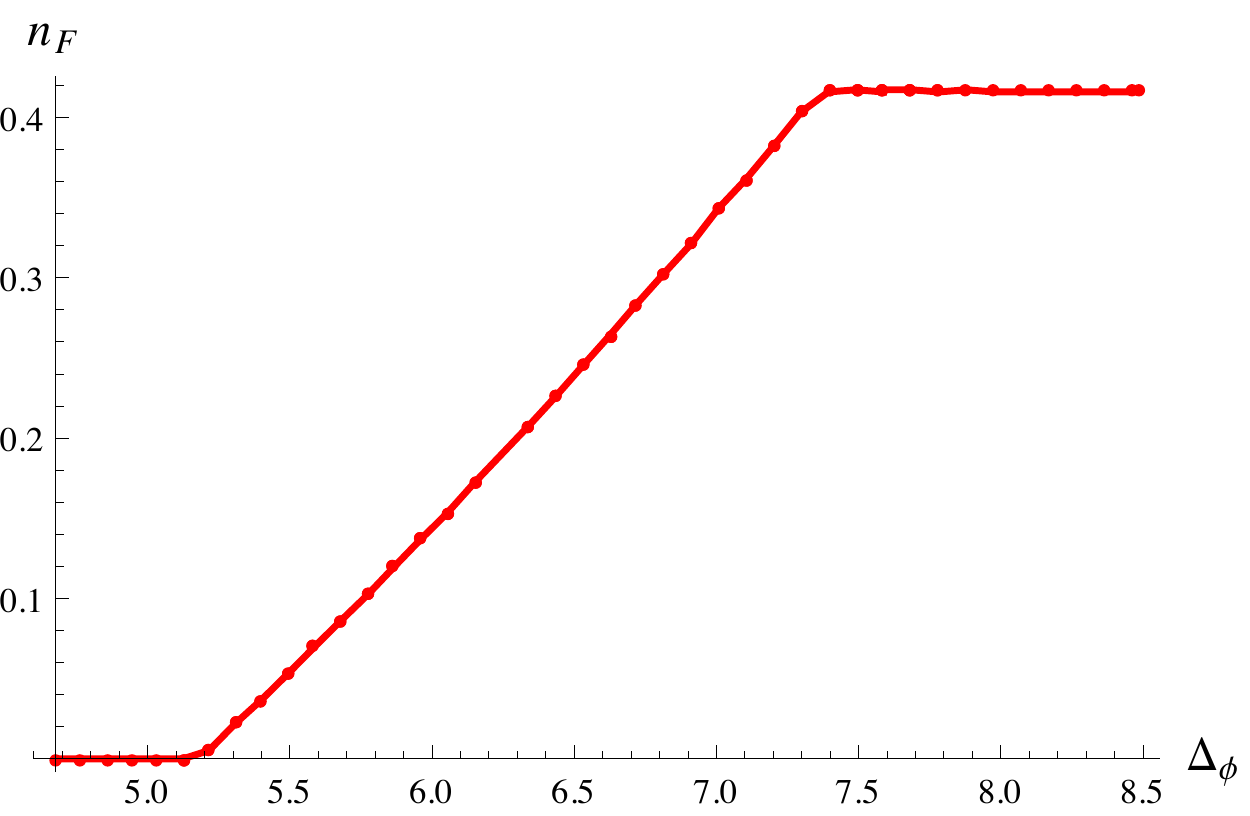}}

\caption{(a): Comparision of the superconducting phase transition in a purely scalar system (blue curve) to the one in a system with fermions at $\eta_5=0$ (red curve). At small conformal
dimension there is no difference between the phase curves at all,
while for larger dimension we see effects of Bose-Fermi competition. (b): Total fermionic bulk charge as a function of
scalar conformal dimension, $n_F=\int\limits_{0}^{z_w}qz^2\langle\psi^\dagger\psi\rangle dz$. Here $\mu q=4.5$, $z_w=1$,
$q_\phi=2$, $\omega_D=0.7$. }
\label{fig:BoseFermi}
\end{center}
\end{figure}

The next step is to see what happens in a system
where both scalar and fermionic fields are present, but interact
with each other only via the gauge field $A_0$, and not directly
(the Majorana coupling $\eta_5=0$ vanishes). For the same parameters
as in Fig. \ref{fig:PureHS} for a scaling dimension of the fermionic
operator $\Delta_\Psi=m_\Psi+3/2 =5/2$ we obtain a scalar condensate
shown on Fig.\ref{fig:BoseFermi}.

Comparing, the two condensate values become identical with the
pure hardwall superconductor without fermions for low enough
$\Delta_{\phi}$. For these values the
bulk scalar field is so light that it consumes all the energy in the system. Ceteris paribus we would need a higher chemical
 potential
 to make fermions occupy the first band and backreact on $A_0$.

At larger values of $\Delta_{\phi}$ there is still a scalar
condensate, but it is suppressed compared to the pure hardwall
superconductor (Fig.\ref{fig:BoseFermi}(a)). This can be easily understood in terms
of canonical ensemble. For fixed total electromagnetic charge of the
system, adding new constituents (fermions) would redistribute the
available charge (Fig.\ref{fig:BoseFermi}(b)) and the condensate of the original degrees of freedom would be suppressed.
This effect has also been observed in a holographic set-up
  where the fermions are approximated in the fluid \cite{Liu:2013yaa,Nitti:2013xaa}

\subsection{A dynamical BCS scalar and a BCS/BEC crossover}
\label{sec:fully-inter-syst}

Now we analyze the most interesting case and include the full dynamics
for the scalar field $\phi$. Let us give another reason why this is
quite natural from the field theory perspective. The evolution in the
radial direction in AdS captures the (leading matrix large $N$
contribution to the) RG flow of the corresponding
operator in the field theory. The BCS gap, proportional to the vev of
the scalar field is certainly sensitive to the RG scale. Hence one
expects it to change dynamically as a function of the radial
direction. Strictly speaking the double trace pairing operator which
sets the value of the gap is a subleading operator in large $N$ and
any running that deviates from its semiclassical scaling is therefore a quantum effect in the AdS gravity
theory. This is the situation we studied in section
\ref{doublegap}. At the $1/N$ level, for small enough coherence
length, the pair operator will become dynamical and qualitatively it
ought to be given by the dynamical scalar we study here.

We will see a very interesting effect occurs in doing so. Because the
scalar is sourced by the Cooper pair condensate, this changes near-boundary
fall off of $\phi$, and the standard holographic prescription for
boundary field theory condensates has to be modified.
Without the presence of a Cooper pair condensate, the zero momentum
scalar mode equation in $AdS_4$ is a homogeneous (linear) differential equation
\be z^2\phi''(z)-2z\phi'(z)+q_\phi^2
z^2A_0^2(z)\phi(z)-m_\phi^2\phi(z)=0.
\ee
Near the AdS boundary its solutions have the following form
\begin{align}
\label{eq:7}
 \phi(z)&=Az^{3-\Delta_\phi}\cdot
\left(1+a_1z+a_2z^2+...\right)+Bz^{\Delta_\phi}\cdot
\left(1+b_1z+b_2z^2+...\right), \nonumber \\
 \Delta_\phi&=\frac{3}{2}+\frac{1}{2}\sqrt{9+4m_\phi^2},
\end{align}
and in the standard quantization scheme the coefficient $A$ of the
non-normalizable solution corresponds to the source $J_{\cO_{\phi}}$ for the operator
$\cO_{\phi}$ dual to $\phi$, and the coefficient $B$ of the
normalizable solution to the vev $\langle
\cO_{\phi}\rangle$. Spontaneous symmetry breaking due to a
condensation of the operator occurs for a
solution in the absence of a source, i.e. with $A=0$ as a boundary condition.

For the interacting scalar-fermion system this simple one-to-one
correspondence between bulk asymptotics and boundary condensates
needs modification. We must now consider the inhomogeneous
differential equation
\be
z^2\phi''(z)-2z\phi'(z)+q_\phi^2 z^2A_0^2(z)\phi(z)-m_\phi^2\phi(z)=-i\eta_{5}z^{3}\langle\overline{\psi^{c}}\Gamma^{5}\psi\rangle .
\ee
The solutions to this equation now include the particular solution
responding to the inhomogeneous source in
addition to the homogeneous solutions \eqref{eq:7}. For near boundary
behavior of the source
\begin{align}
\label{eq:8}
\lim\limits_{z\rightarrow
  0}z^{3}\langle\overline{\psi^{c}}\Gamma^{5}\psi\rangle
\sim z^{2\Delta_{\Psi}}
\end{align}
the particular solution will behave in the same way (assuming ${2\Delta_{\Psi}}\neq {\Delta_\phi}$):
\bea
\phi(z)& =&\phi_{hom}(z)+\phi_{part}(z) \nonumber\\
 \phi_{part}(z) &=& \cP_1 z^{2\Delta_{\Psi}} + \cP_2 z^{2\Delta_{\Psi}+1}+\cP_3z^{2\Delta_{\Psi}+2}+... \label{InhomSeries}
\eea

This particular solution will control the dominant normalizable near
boundary behavior for $\Delta_{\phi}>2\Delta_{\Psi}$.
 This raises the question what we should use as the vev
for the corresponding operator. The canonical AdS/CFT prescription
\begin{align}
  \label{eq:9}
  \langle \cO_{\phi} \rangle = \lim_{z\rar 0}
  z^{-d+1} \partial_z \left(z^{d-\Delta_{\phi}} \phi(z)\right)
\end{align}
no longer gives a viable answer. Let us exhibit this in detail.
As an aside, note that the near-boundary behavior of the fermions does not change
provided the solution for $\phi(z)$ is normalizable.

\begin{figure}[t!]
\begin{center}
\includegraphics[scale=0.6]{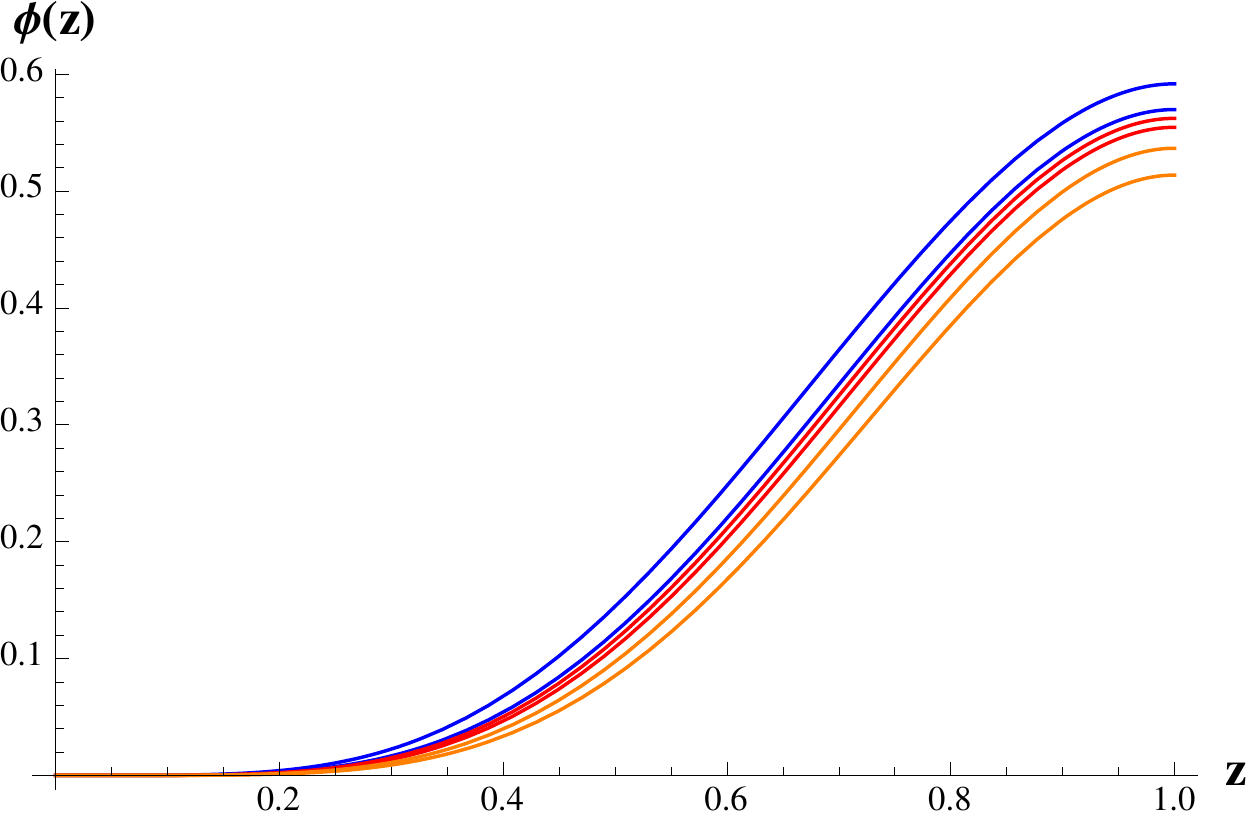}
\caption{Profiles of the bulk scalar wavefunction $\phi(z)$ for $\Delta_\phi=4.6765$, $\Delta_\phi=4.8541$ (two blue curves), $\Delta_\phi=4.9438$, $\Delta_\phi=5.0341$ (two red curves, - proximity of the critical point), $\Delta_\phi=5.379$, and $\Delta_\phi=5.4925$ (two orange curves). Crossing
the critical point $\Delta_\phi=2\Delta_{\Psi}=5$ does not lead to any singularities in the bulk wavefunction. The other parameters
here are $\eta_5=1,\,\mu q=4.5,\,\mbox{$z_w=1$},\,q_\phi=2,\,\omega_D=0.7$.}
\label{fig:RegularWF}
\end{center}
\end{figure}

\begin{figure}
\begin{center}
\includegraphics[width=0.5\textwidth]{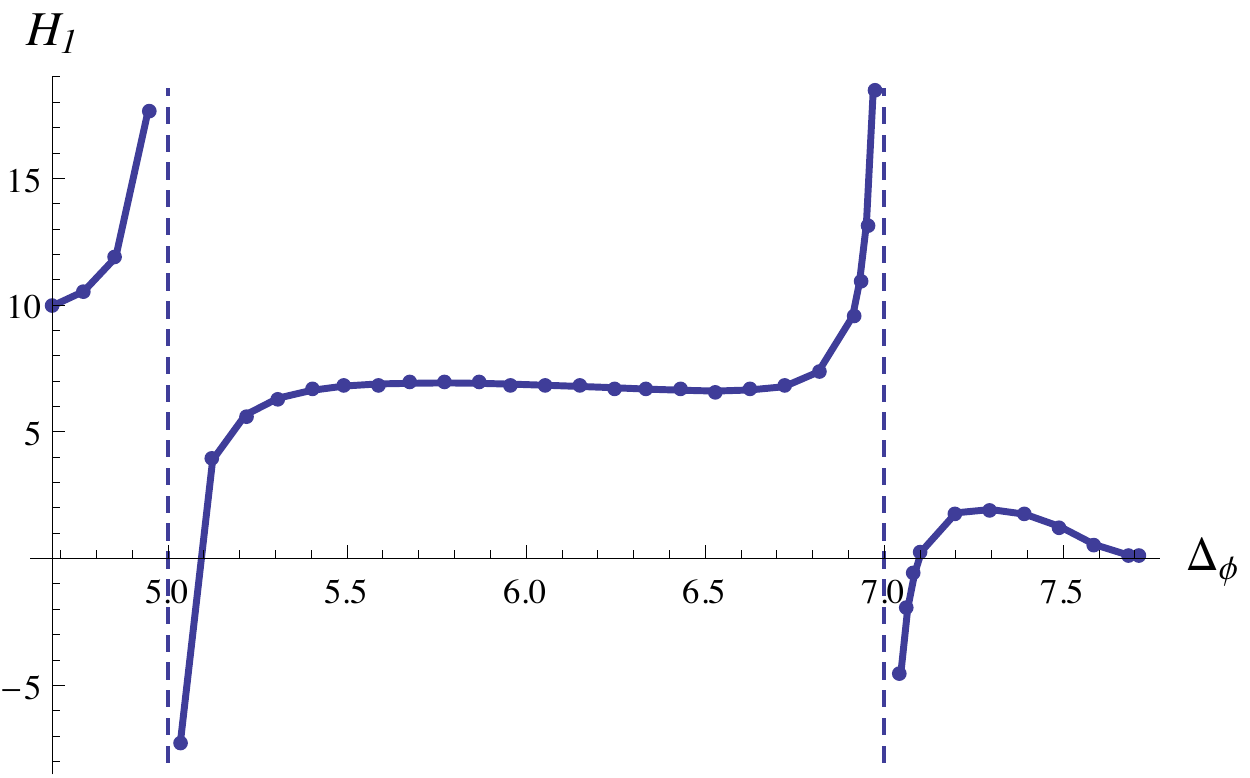}\caption{Dependence of the leading homogeneous coefficient in the scalar solution expansion on the conformal dimension of the field.
Here $\mu q=4.5,\,\eta_5=1,\,z_w=1,\,q_\phi=2,\,m_\Psi=1$ (so $2\Delta_{\Psi}=5$),\,$\omega_D=0.7$.}
\label{fig:resonances1}
\end{center}
\begin{center}
\begin{tabular}{cc}

\includegraphics[width=0.49\textwidth]{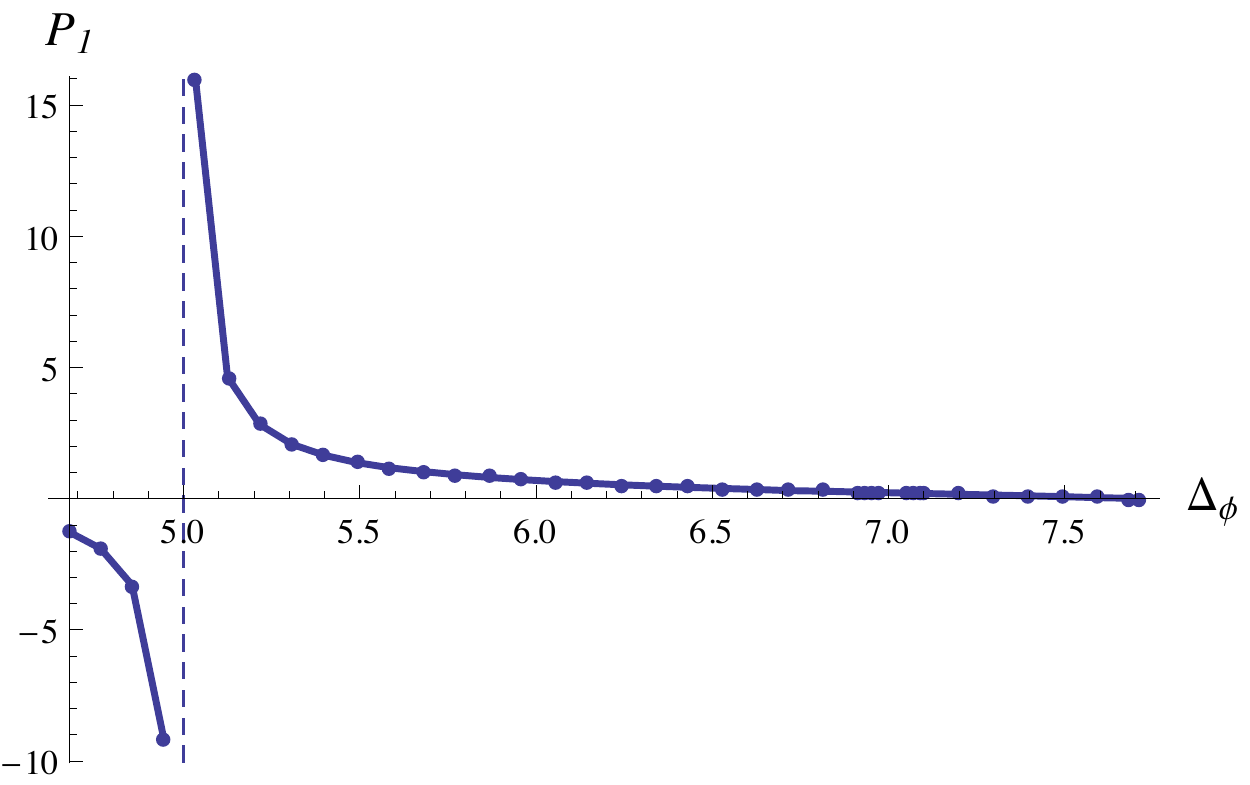}
\includegraphics[width=0.49\textwidth]{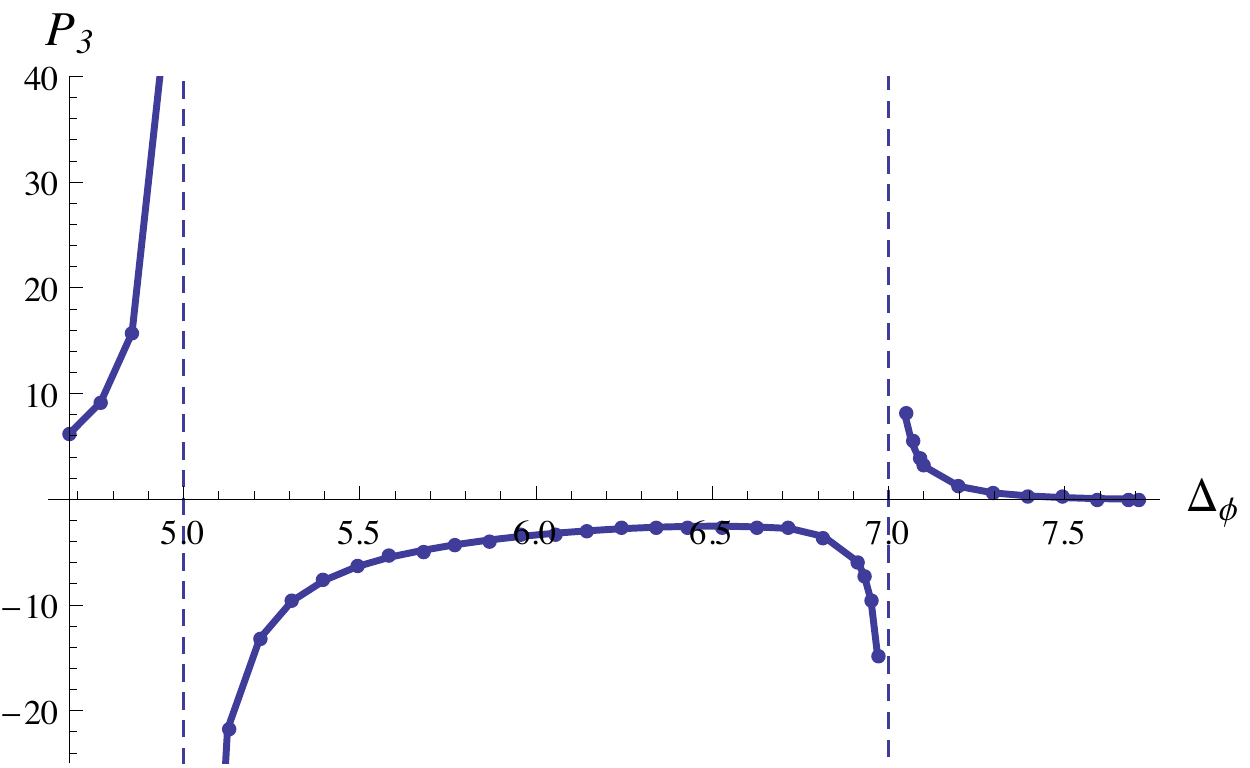}

\end{tabular}
\caption{Dependence of the two leading particular coefficients in the scalar solution expansion on the conformal dimension of the field.
Here $\mu q=4.5,\,\eta_5=1,\,z_w=1,\,q_\phi=2,\,m_\Psi=1$ (so $2\Delta_{\Psi}=5$),\,$\omega_D=0.7$. }
\label{fig:resonances2}
\end{center}
\end{figure}

Denoting the coefficient $B$ of the normalizable homogeneous solution
with $B=\cH_1$ we extract these coefficients from numerical solutions to the scalar and fermionic equations.
(see Fig.\ref{fig:resonances1}, \ref{fig:resonances2}). Immediately
noticable are the singularities at $\Delta_{\phi}=2\Delta_{\Psi}$
and $\Delta_{\phi}=2\Delta_{\Psi}+2$.
Strictly speaking when $\Delta_{\phi}=2\Delta_{\Psi}+n$ the expansion
\eqref{InhomSeries} breaks down and the
solution has an extra logarithmic term
\be
\phi(z)=\cH_1 z^{2\Delta_{\Psi}+n}+...+\cP_1 z^{2\Delta_\Psi}+...+ \cP_{n+1} z^{2\Delta_\Psi+n}\ln(z)+...
 \label{InhomSeries2}
\ee
The singular divergence of coefficients is a precursor of this
logarithm.
 There is no singularity at
$2\Delta_{\Psi}+1$ because $\cP_2$ happens to vanish in our case.\footnote{This vanishing of ${\cal P}_2$ (due to the vanishing of ${\cal S}_2$) and the structure of the series expansion is determined by the solutions of the Dirac equation. For zero electric field each {\it even } coefficient would
vanish in fact. Since the gauge field profile modifies the higher order coefficients in the series expansion of the
Dirac equation, it can be shown that ${\cal S}_4\neq0$.}

The indisputable presence of these singularities or resonances can be
readily seen by considering a simplified version of the scalar
equation. Computing the series solution to the equation
\begin{align}
  \label{eq:10}
  z^2\phi'' -2 z\phi' - m_{\phi}^2 \phi = \cS_1 z^{2\Delta_{\Psi}} + \cS_3 z^{2\Delta_{\Psi}+2}
\end{align}
one directly finds the ``resonances''
\begin{align}
{\cal P}_1&=\frac{{\cal
    S}_1}{2\Delta_{\Psi}(2\Delta_{\Psi}-3)-\Delta_\phi(\Delta_\phi-3)},
\nonumber \\
{\cal P}_3&=
\frac{{\cal
    S}_3}{(2\Delta_{\Psi}+2)(2\Delta_{\Psi}-1)-\Delta_\phi(\Delta_\phi-3)}.
\label{eq:AlgebraicResonance}
\end{align}
Note that they are Feschbach-like resonances in that the singularity is a single rather than a double pole.

\begin{figure}[t!]
\begin{center}
\begin{tabular}{cc}
\includegraphics[width=0.5\textwidth]{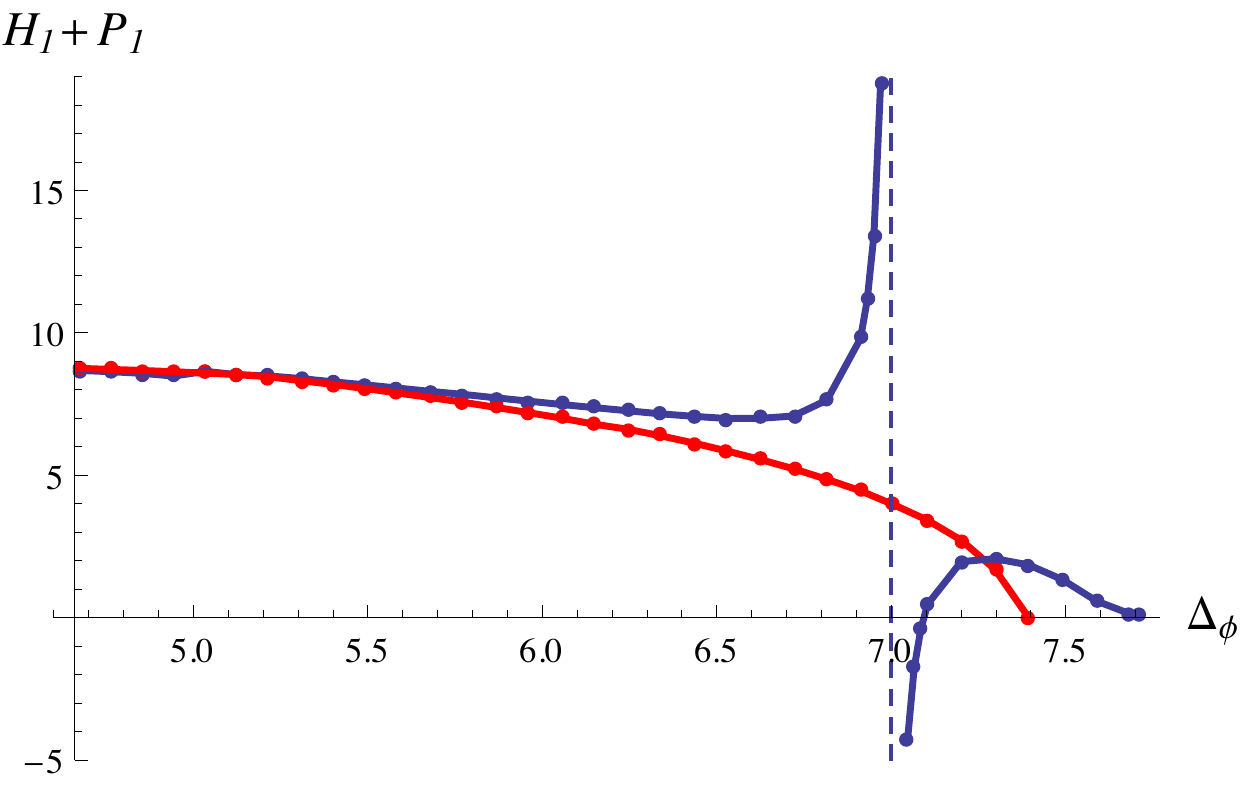}
\includegraphics[width=0.5\textwidth]{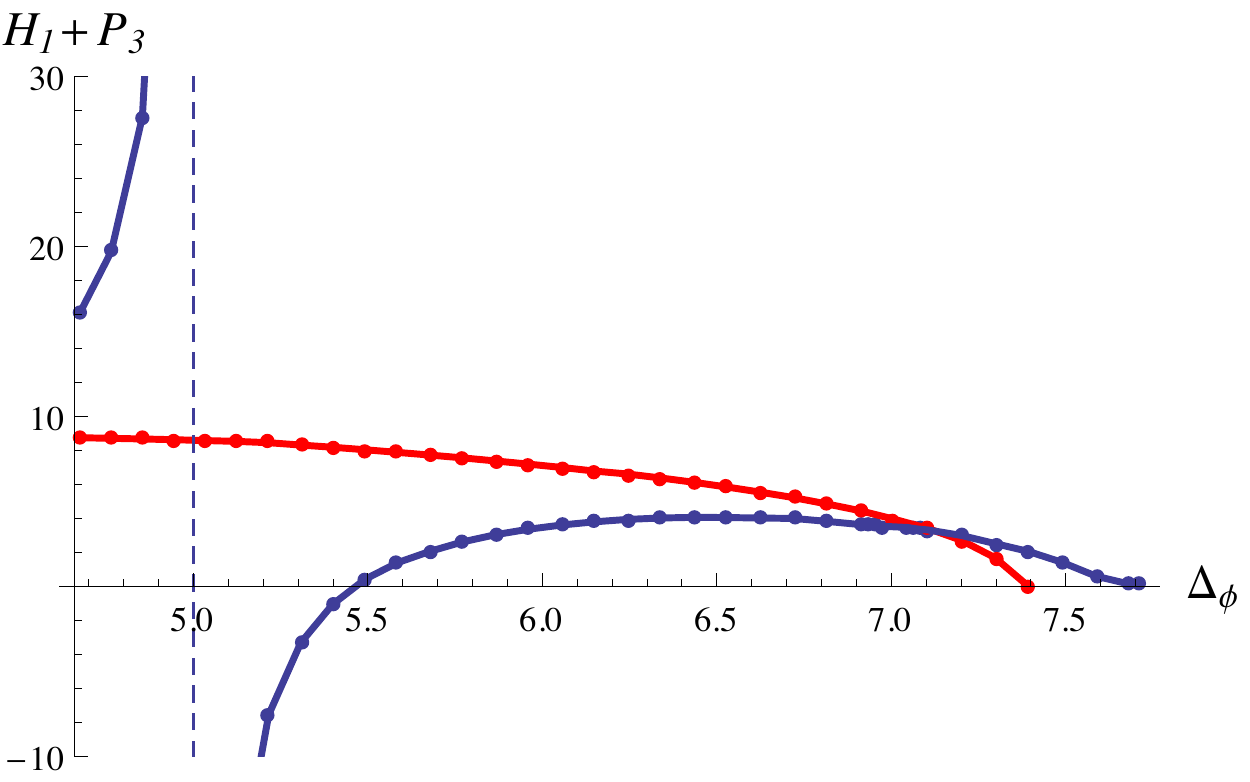}
\end{tabular}
\caption{Linear combinations of the series expansion coefficients. Red curve represents the boundary scalar operator condensate $\langle \cO_\phi \rangle$ as a function of its conformal dimension in presence of fermions at $\eta_5=0$ (the same as the red curve on Fig.\ref{fig:BoseFermi}). The blue curve represent the linear combination ${\cal H}_1+{\cal P}_1$ on the left plot, and ${\cal H}_1+{\cal P}_3$ on the right one. Resonances in ${\cal H}_1$ and ${\cal P}_1$ precisely cancel
each other at $\Delta_\phi=2\Delta_{\Psi}$, and so do resonances in ${\cal H}_1$ and ${\cal P}_3$ at $\Delta_\phi=2\Delta_{\Psi}+2$. All parameters are as in Fig.\ref{fig:resonances1}.}
\label{fig:canceled-resonances}
\end{center}
\end{figure}

The question is how to extract the information of the strongly
coupled dual field theory from this asymptotic behavior of the AdS scalar
wavefunction. Despite these singularities in the coefficients, by
construction the bulk scalar wavefunction is regular at all points
(Fig.\ref{fig:RegularWF}). It is therefore physically natural to have
regular observables in the boundary field theory as well. There are
two obvious points to make here.
\begin{itemize}
\item[(1)] Physically the origin of the order parameter is
  indistinguishable. One cannot tell whether the broken groundstate is
  caused by condensation of the Cooper pair or the scalar field.
\item[(2)] Mathematically, the regularity of the bulk solution
  directly implies that the homogeneous component $\cH_1$ must have a
  similar resonance but with an opposite sign.
\end{itemize}
An obvious and physically motivated choice is to postulate that the
actual order parameter is the simply the sum of the naive condensates,
with the Cooper pair condensate $\cS_1$ renormalized to $\cP_1$: i.e.
\be \langle
{\cal O}_\phi \rangle = {\cal H}_1+{\cal P}_1.
\ee
Taking this linear combination does in fact lead to a
cancelation of ``resonances'' and a smooth function  at
$\Delta_\phi=2\Delta_{\Psi}$ (see Fig.
\ref{fig:canceled-resonances}). However, the
reflection of the next resonance
$\Delta_\phi=2\Delta_{\Psi}+2$ in the homogenous solution $\cH_1$ remains.
Likewise, a similar partial resolution occurs for the linear
combination $\cH_1+\cP_3$.

\begin{figure}[t!]
\begin{center}
\begin{tabular}{cc}
\subfigure[]{\includegraphics[width=0.5\textwidth]{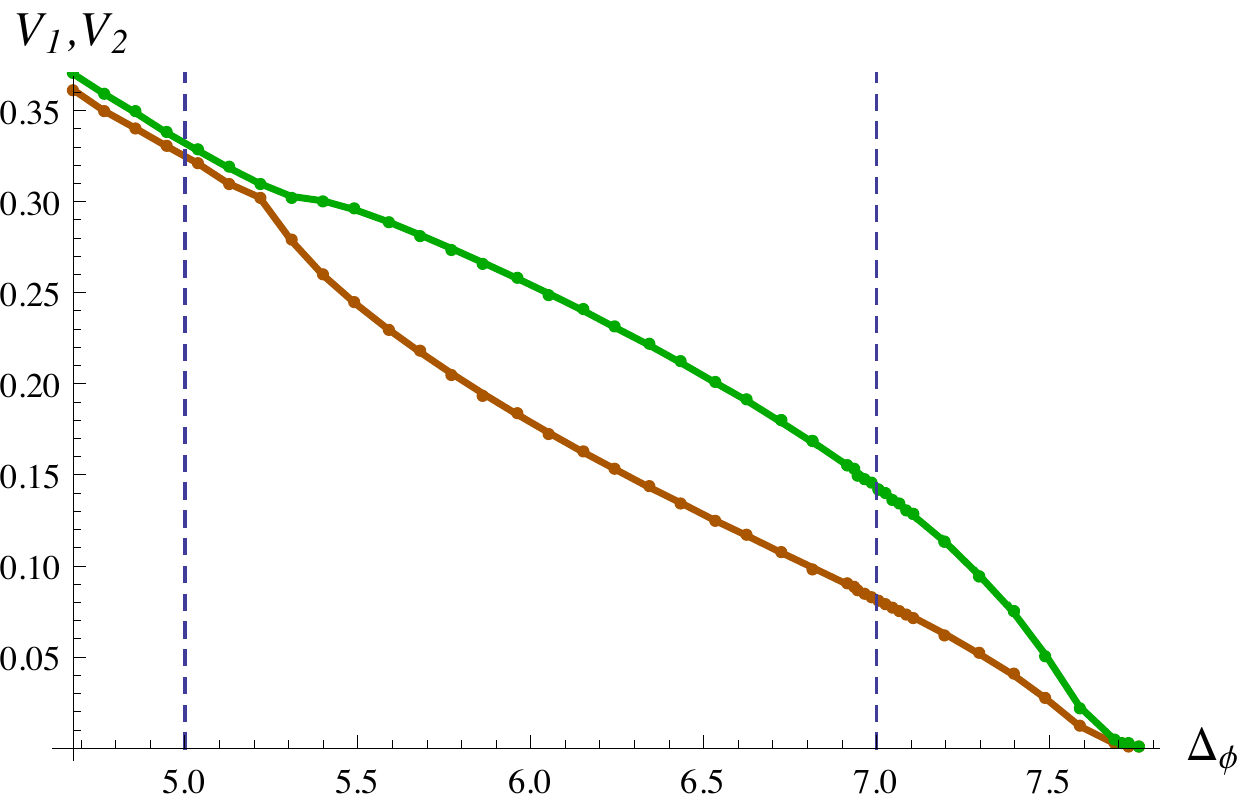}}
\subfigure[]{\includegraphics[width=0.5\textwidth]{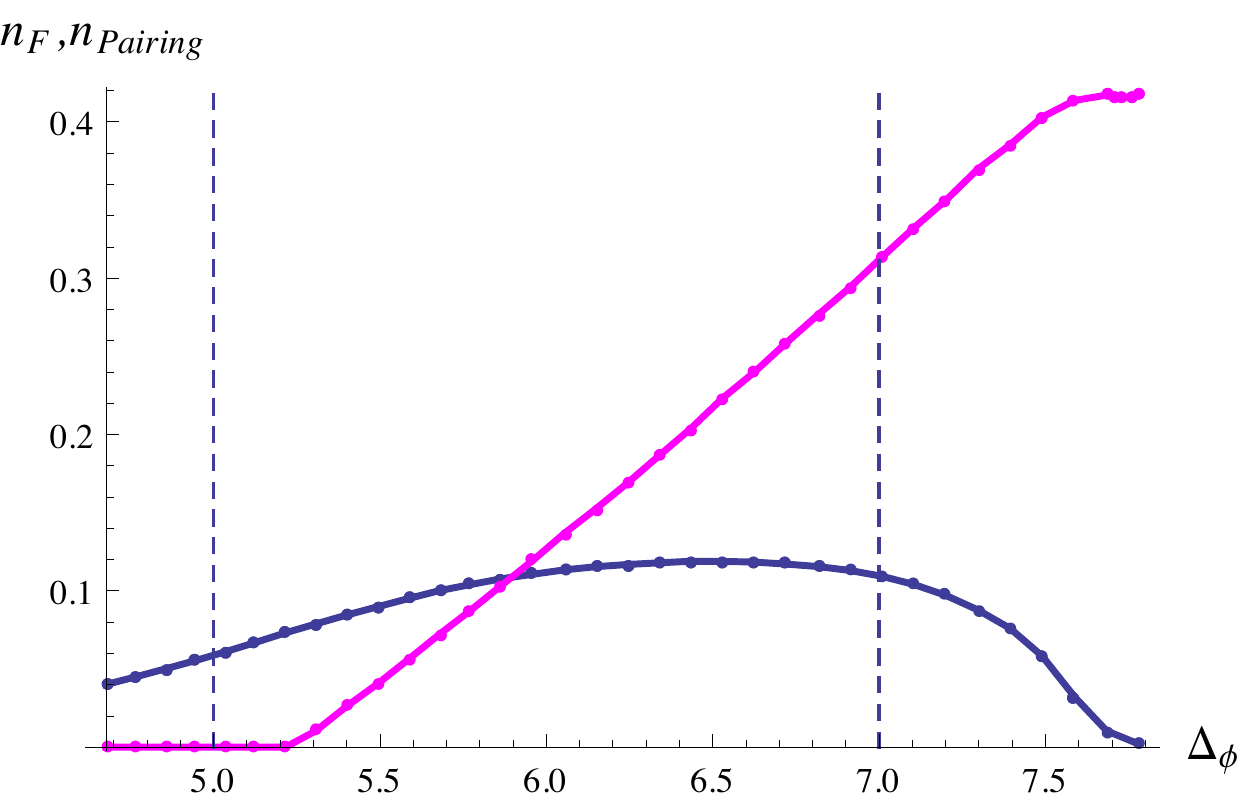}}
\end{tabular}
\caption{(a): Dependence of the two gaps $V_1$ (orange) and $V_2$ (green) on $\Delta_\phi$ in the fully interacting case. (b): Dependence of the total fermionic bulk charge $n_F=\int\limits_{0}^{z_w}qz^2\langle\psi^\dagger\psi\rangle dz$ (magenta) and the total ``number'' of pairs $n_{Pairing}=-i\eta_5\int\limits_{0}^{z_w}z^3\langle\overline{\psi^C}\Gamma^5\psi\rangle dz$ (blue) on $\Delta_\phi$.
One can see that while at small scalar conformal dimensions the fermionic bulk charge totally vanishes the number of Cooper pairs in the bulk theory stays finite.
All parameters are as in Fig.\ref{fig:resonances1}.}
\label{fig:GapsAndCharges}
\end{center}
\end{figure}

\begin{figure}[t!]
\begin{center}
\begin{tabular}{cc}
\subfigure[]{\includegraphics[width=0.5\textwidth]{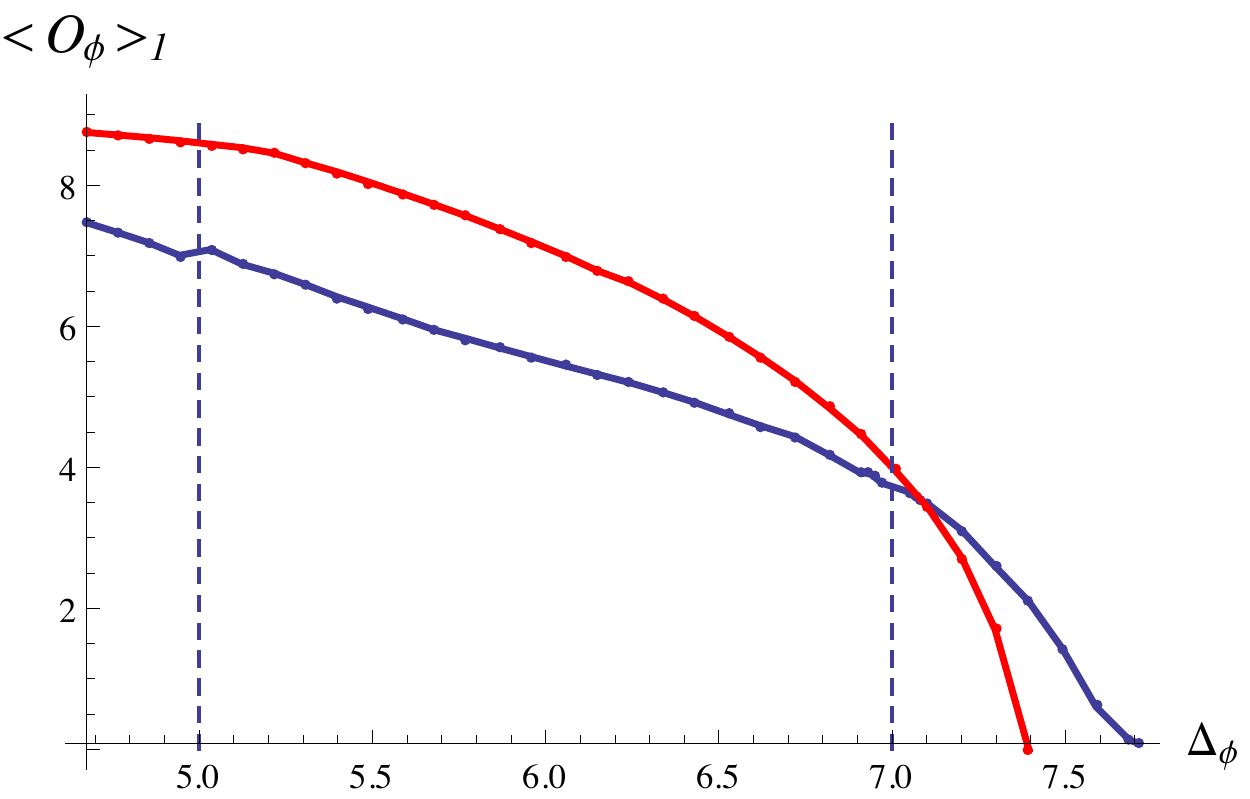}}
\subfigure[]{\includegraphics[width=0.5\textwidth]{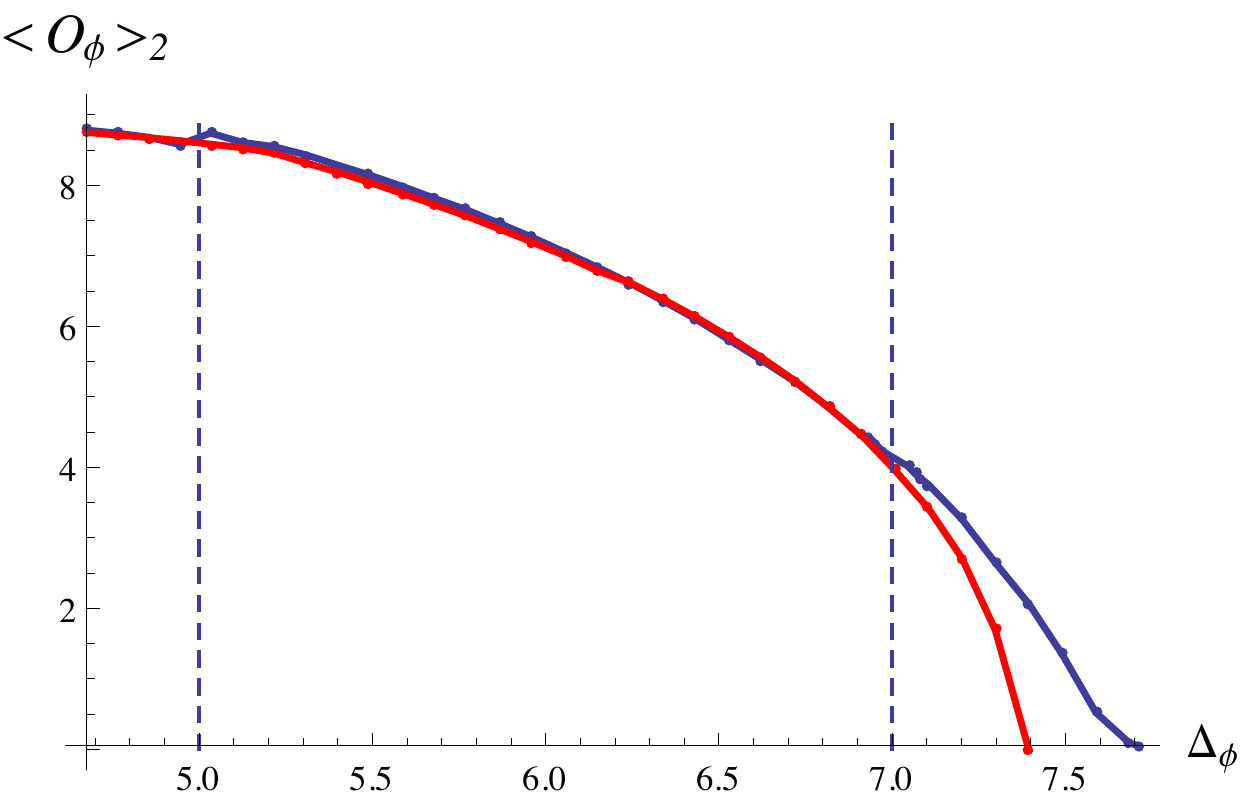}}
\end{tabular}
\caption{(a): The blue curve represents a particular linear combinations of the series expansion
  coefficients $\langle\cO_\phi\rangle_1 = \cH_1 + \frac{1}{2}((2\Delta_{\Psi}+2)-\Delta_{\phi}
  )\cP_1 + \frac{1}{2}(\Delta_{\phi}-2\Delta_{\Psi})\cP_3$ such that
  all the resonances cancel out. The red
  curve included for comparison represents $\langle\cO_\phi\rangle$ at
  $\eta_5=0$. (b): The serendipitous combination
 $\langle\cO_\phi\rangle_2 = \cH_1 + \frac{1}{2}e^{-(2\Delta_{\Psi}-\Delta_\phi)}((2\Delta_{\Psi}+2)-\Delta_{\phi}
  )\cP_1 +
  \frac{1}{2}e^{-(2\Delta_{\Psi}+2-\Delta_\phi)}(\Delta_{\phi}-2\Delta_{\Psi})\cP_3$
  that has a remarkable overlap with the $\eta_5=0$ solution at low
  $\Delta_{\phi}$ as desired.
  All parameters are as in Fig.\ref{fig:resonances1}.}
\label{fig:lincomb}
\end{center}
\end{figure}

These ``resonances'' in $\cP_i$ and their cancellation by (part of)
the homogeneous solution $\cH_1$ will in fact occur at every order of
the expansion from the AdS boundary $z=0$. It hints that the proper definition of the superconducting
condensate should be given by a $\Delta_\phi$-dependent
linear combination of the homogeneous ${\cal H}_1$ and
particular coefficients ${\cal P}_1,{\cal P}_3,{\cal
P}_5,...$ that is regular for all $\Delta_{\phi}$. One can readily construct such a combination, e.g.
\begin{align}
  \label{eq:12}
\langle\cO_\phi\rangle = \cH_1 + \frac{1}{2}((2\Delta_{\Psi}+2)-\Delta_{\phi}
  )\cP_1 + \frac{1}{2}(\Delta_{\phi}-2\Delta_{\Psi})\cP_3,
\end{align}
see Fig. \ref{fig:lincomb}(a).
as a demonstration of the existence of a non-singular
combination; though there is no proof at all that this constitutes
the actual physical observable.\footnote{Another putative combination found by chance, $\langle\cO_\phi\rangle = \cH_1 + \frac{1}{2}e^{-(2\Delta_{\Psi}-\Delta_\phi)}((2\Delta_{\Psi}+2)-\Delta_{\phi}
  )\cP_1 +
  \frac{1}{2}e^{-(2\Delta_{\Psi}+2-\Delta_\phi)}(\Delta_{\phi}-2\Delta_{\Psi})\cP_3$
  has a remarkable overlap with the scalar condensate in the case
  $\eta_5=0$, see Fig. \ref{fig:lincomb}(b).}

Normally the strict application of the AdS/CFT dictionary does
not assign any role to such higher order coefficients in the bulk
wavefunction. It is clear, however, that the singularities arise solely from the extraction of the coefficients, whereas the full AdS wavefunctions at any finite $z$ are regular for $\Delta_{\phi}=2\Delta_{\Psi}+\mathbb{N}$. Let us now give an argument why the coefficient rule can receive modification.The right way to interpret the
linear combination ${\cal H}_1+{\cal P}_1$ is as a mixing of the two independent
operators dual to the fundamental scalar operator and the bilinear
(double trace)
Cooper pair operator. This suggests that we should think in a similar way
about the resonance at $\Delta_{\phi}=2\Delta_{\Psi}+2$.
There should be another
Cooper-pair like operator in the theory which mixes with the
fundamental scalar, such that the linear combination that constitutes
the order parameter is finite.

In AdS/CFT this connection between mixing and resonances is in fact cleanly seen in
correlation functions of bilinear operators
\cite{Fitzpatrick:2010zm,Fan:2011wm}. These bilinear operators are also known as
double trace operators, since in the models where we know the dual
CFT, each operator dual to an
AdS field is a single trace over an $N\times N$ matrix valued
combination of fields. Bilinear operators are thus the normal-ordered
product of two single trace operators. Each pair of single trace CFT operators ${\cal O}_\Psi$,
however, gives rise to an infinite tower of {\em independent primary} double trace
operators:
\newcommand\lrpar{\raise .8ex\hbox{$^\leftrightarrow$} \hspace{-9pt}
\partial}
\newcommand\lpar{\raise .8ex\hbox{$^\leftarrow$} \hspace{-9pt}
\partial}
\newcommand\rpar{\raise .8ex\hbox{$^\rightarrow$} \hspace{-9pt}
\partial}
\begin{align}
\cO_{(0)} &= {\cal O}_{\overline{\Psi^C}}{\cal O}_\Psi  \nonumber\\
\cO_{(1)} &={\cal O}_{\overline{\Psi^C}}(\lpar_{\mu}-\rpar_{\mu})(\lpar^{\mu}-\rpar^{\mu})
{\cal O}_\Psi - \mathrm{trace}
\nonumber\\
\cO_{(2)} &={\cal O}_{\overline{\Psi^C}}
(\lpar_{\mu}-\rpar_{\mu})
(\lpar^{\mu}-\rpar^{\mu})
(\lpar_{\nu}-\rpar_{\nu})
(\lpar^{\nu}-\rpar^{\nu})
{\cal O}_\Psi - \mathrm{traces}
\nonumber\\
&\vdots
\end{align}
These \comment{CHECK} {\em conformal partial waves} are all the higher
derivative bilinear operators that cannot be written as a descendant (a
derivative) of the a lower order primary.
All these operators have the same global quantum numbers as the simple pair
 operator with scaling dimension $2\Delta_{\Psi}$, but increase their
 dimension by two integer units each time.
 The correlation function study \cite{Fitzpatrick:2010zm,Fan:2011wm}
 in particular shows that in the case of an interacting  purely
scalar bulk theory, {\em all} these linearly independent
 double trace
primaries mix in as well and cause single-pole Feschbach resonances in s-wave scattering of single trace operators. The correspondence between the $2n$ difference in scaling dimension\footnote{As we mentioned one also expects a resonance at $2\Delta_{\Psi}+3$ for high enough chemical potential. This is due to the effect of the electric field on the fermion wave functions. From the boundary perspective this could be a result of mixing with ${\cal O}_\psi J_{\mu}(\lpar^{\mu}-\rpar^{\mu})
{\cal O}_\psi$ type operator which has the right scaling dimension ($\Delta_J=2$).}
 between each successive primary and the location of the resonance in
 the leading part of the bulk scalar wavefunction supports that this
 mixing is the right interpretation of the resonance.\footnote{The conformal partial wave operators share a resemblance
with operators relevant for Fulde-Ferrel-Larkin-Ovchinnikov pairing
\cite{Ful64,Lar64}. In the original FFLO set-up one considers
the Zeeman splitting of spin-up/spin-down electrons and this causes
an offset in their Fermi surfaces of the same form seen here.
The discussion about the mixing in of these higher order partial waves
does not rely on the split degeneracy of Fermi surfaces. The mixing is
therefore not correlated with an FFLO-like phenomenon.}

We do not yet have a controlled method to extract the quantative expectation
 value of these higher order double trace primaries from the
 constituent single trace fields. The mixing originates in the
 renormalization of the theory, and this suggests that the proper
 value of the order parameter results from the introduction of higher
 order boundary counterterms of the type
 \begin{align}
   \label{eq:13}
   S_{counter} \sim \int_{z=\eps} d^3 x\left( - \phi^2 - \phi
   \bar{\Psi}^C_+\Psi_- - \phi
   \bar{\Psi}^C_+\lrpar_{\mu}\lrpar^{\mu}\Psi_- - \ldots \right)
 \end{align}
where $\Psi_{\pm}$ are eigenspinors of $\Gamma^5$. To construct this
correct set of counterterms and deduce the appropriate extraction of
the vev in the boundary field theory is an interesting question to
pursue.

The conclusion is that the resulting condensate ought to be of the
form in Fig. \ref{fig:lincomb}. Qualitatively this result shows the
BCS/BEC crossover as a function of the scalar scaling dimension $\Delta_{\phi}$. Though our set up is
rather abstract in that scalar field here is an
additional degree of freedom introduced by hand, instead of emerging
from microscopic dynamics, it
captures the BCS/BEC physics.
For small scaling dimension the
scalar operator ${\cal O}_\phi$ dominates the Bose-Fermi
competition, whereas
at
large scalar conformal dimensions corresponding to weak coupling
regime, $\eta_5/m_\phi\ll 1$, the dynamics of the boson field are suppressed,
and its order parameter expectation value is dominated by fermions
as shown on Fig. \ref{fig:canceled-resonances}.
The most interesting region is just to the right of the red
curve. Here there is no bosonic contribution to the order parameter,
but there is an enhanced Cooper pair contribution (due to the
proximity effect). This is the most notable region where we have
pairing induced superconductivity in holography.
At larger scalar conformal dimension the order parameter exponentially
decreases with increasing of $\Delta_\phi$, although it never
vanishes. In the strict $m_{\phi} \rar \infty$ limit we have the
standard BCS scenario of section \ref{doublegap}.

Let us finally briefly comment on the
dependence on the UV cut off $\omega_D$. In the previous section we discussed that at very large bulk scalar mass
all dynamics depends only on two parameters, $\eta_5/\omega_D$ and
$\eta_5/m_\phi$. For a dynamical scalar the dependence is more
complicated, but we can still qualitatively infer what will happen. We
know that most of the contribution to the pairing operator is located
near the Fermi surfaces. Increasing $\omega_D$ means taking into
account states lying far away from $k_F$'s. The physical picture will
therefore only change minimally; to first approximation it can be
compensated by adjusting $\eta_5$ such that $\eta_5/\omega_D$ stays
constant. A non-trivial effect does happen when $\omega_D$ becomes so
large that the integral becomes sensitive to fermions in the second band (for instance, see Fig. \ref{fig:plasmino}), but this is beyond the scope of this paper.

\section{Conclusions}

We have constructed a holographic model of superconductivity which explicitly takes into account
fermionic pairing driving the phase transition. In the simplest holographic models, the microscopic mechanism
of superconductivity is not addressed. Specific top-down models may shed light on the strong coupling dynamics and
a possible pairing mechanism \cite{Ammon:2009fe,Ammon:2010pg}, but generic holographic models operate at a Landau-Ginzburg order parameter level.

Even so, the physics of fermionic pairing and condensation should also be explicitly representable in holographic systems.
The most straightforward way to do so is to mimic the classic BCS mechanism. This is what we have done here. By introducing
an attractive four-fermion interaction in the $AdS$ bulk, we show that this directly reduces to a pairing induced superconducting groundstate both in the bulk and the dual boundary. To cleanly separate the fermion physics, we introduced
a hard wall cut-off. This essentially guaranteed this results as the low energy theory in both sides is just a Fermi liquid
in the absence of the four-fermion interaction. The one technical difference with textbook BCS is the relativistic nature
of the underlying fermion theory.

Next we introduced separately a kinetic term for the $AdS$ dual of the order parameter. Physically the paired operator should
become dynamical if the coherence length is much shorter than the scales of interest. One should find a BCS/BEC
crossover as one tunes between these regimes. Here that control parameter is the scaling dimension of the order parameter
field (relative to the scaling dimension of the Cooper pair operator). For large scaling dimension
the kinetics of the dual $AdS$ field is suppressed and we have the BCS physics found
earlier. For low scaling dimension the scalar dynamics should be energetically favored compared to pairing condensation, and one should find a regular BEC (holographic) superconductor.

In observing this BCS/BEC crossover we encountered a surprise. At specific values $\Delta_\phi=2\Delta_\Psi$ and
$\Delta_\phi=2\Delta_\Psi+2$ of the control parameter the independent scalar $\langle \cO_\phi\rangle$ and pairing
$\langle\cO_{\overline{\Psi^C}}\cO_{\Psi}\rangle$ vevs diverge. In fact the naive order parameter
$\langle \cO_\phi\rangle+\langle\cO_{\overline{\Psi^C}}\cO_{\Psi}\rangle$ remains divergent at $\Delta_\phi=2\Delta_\Psi+2$.
The mathematics is clear and suggests that these divergences can also occur at higher value of the scaling dimension.
Physically, a plausible explanation is that higher order primaries ${\cal O}_{\overline{\Psi^C}}(\lpar_{\mu}-\rpar_{\mu})^n(\lpar^{\mu}-\rpar^{\mu})^n
{\cal O}_\Psi$, that arise in the OPE of the product of two single fermion operators,
mix in with the scalar vev and the lowest order primary $\langle \cO_\phi\rangle+\langle{\cal O}_{\overline{\Psi^C}}\cO_{\Psi}\rangle$.
To establish this concretely requires a more detailed study of single and double trace operator mixing in
$AdS/CFT$. We aim to address this in a future publication. We can
nevertheless readily construct an extraction rule for a finite order parameter that interpolates between
the BCS and BEC regimes.

In both aspects the physics that holographic system describes is very conventional. It is again an excellent proving ground
for $AdS/CFT$ that it does so, but by construction it does not uncover any unconventional or exotic physics. The main reason it
does not do so is the presence of the hard wall. It ensures that the groundstate dynamics closely follows standard Fermi liquid and Landau-Ginzburg theory. It would be very interesting, but technically challenging \cite{Allais:2012ye,Allais:2013lha},
to try to remove the hard wall. This would reintroduce the low energy dynamics that could yield exotic and novel behaviour. In
particular, it might be an important step towards a holographic fermionic theory of unconventional superconductivity.

\acknowledgments

We are grateful to Irina Aref`eva, Olga Borovkova, Mihailo Cubrovic, Richard
Davison, Sarang Gopalakrishnan, Matthias Kaminski, Subir Sachdev,
Andrei Starinets, Philip Strack, Jan Zaanen and especially Yan Liu and Ya-Wen Sun for discussions. We
are especially thankful to the Condensed Matter Theory Group at
Harvard University for their extensive hospitality. This research
is supported in part by a VICI grant of the Netherlands
Organization for Scientific Research (NWO), by the Netherlands
Organization for Scientific Reseach/Ministry of Science and
Education (NWO/OCW), by a Huygens Fellowship (B. M.), by the
Foundation for Research into Fundamental Matter (FOM), and by
the Russian Foundation for Basic Research (grant 12-01-31298,
A.B.).

\appendix
\section{Green's functions and charge densities} \label{}

In this Appendix we provide a detailed derivation of the formulas
for the fermionic bilinears appearing in the bosonic equations. In
principle while calculating these objects one needs to be careful
because of the renormalization of these composite operators.
However, we are just regularizing these object with a cut off and
not attempting to perform the renormalization. We can write the
fermionic electric charge density as a limit of Feynman Green's
function:
\begin{equation}
\langle\psi^{+}(x)\psi(x)\rangle=\lim_{t,\vec{x} \rightarrow
t',\vec{x}'}\langle
T\psi^{+}(t,\vec{x})\psi(t',\vec{x}')\rangle=\lim_{t,\vec{x}
\rightarrow t',\vec{x}'}G_{\psi_{i}^{+}\psi_{i}}
\end{equation}

We would like to express it with the Nambu-Gorkov (NG) Green's function defined as
\begin{equation}
G_{\chi_{i}\chi_{j}^{+}}=\frac{1}{Z}\int D\chi D\chi^{+}\chi_{i}\chi_{j}^{+}\exp\left(i\int d^{4}x\chi^{+}\widetilde{K}\chi\right).
\end{equation}

Using properties of the time ordered product the relations between
the original Green's functions and the NG ones are
\begin{equation}
G_{\psi_{1}^{+}\psi_{1}}\left(t,\vec{x};t',\vec{x}'\right)=-G_{\chi_{1}\chi_{1}^{+}}\left(t',\vec{x}';t,\vec{x}\right),
\end{equation}
\begin{equation}
G_{\psi_{3}^{+}\psi_{3}}\left(t,\vec{x};t',\vec{x}'\right)=G_{\chi_{3}\chi_{3}^{+}}\left(t,\vec{x};t',\vec{x}'\right).
\end{equation}

With these the charge densities can be expressed with the components of the NG Green's function
\begin{align}
\langle\psi^{+}\psi\rangle= \lim_{t,\vec{x} \rightarrow
  t',\vec{x}'}&\left(-G_{\chi_{1}\chi_{1}^{+}}\left(t',\vec{x}';t,\vec{x}\right)-G_{\chi_{2}\chi_{2}^{+}}\left(t',\vec{x}';t,\vec{x}\right)
\right. \nonumber \\
&\left.
+G_{\chi_{3}\chi_{3}^{+}}\left(t,\vec{x};t',\vec{x}'\right)+G_{\chi_{4}\chi_{4}^{+}}\left(t,\vec{x};t',\vec{x}'\right)\right),
\nonumber
\\
\label{eq:SGreen}
\langle\overline{\psi^{c}}\Gamma^{5}\psi\rangle=
\lim_{t,\vec{x} \rightarrow
  t',\vec{x}'}&\left(G_{\chi_{1}\chi_{4}^{+}}\left(t,\vec{x};t',\vec{x}'\right)+G_{\chi_{2}\chi_{3}^{+}}\left(t,\vec{x};t',\vec{x}'\right)\right.\nonumber\\
&\left.+G_{\chi_{2}\chi_{3}^{+}}\left(t',\vec{x}';t,\vec{x}\right)+G_{\chi_{1}\chi_{4}^{+}}\left(t',\vec{x}';t,\vec{x}\right)\right).
\end{align}

Since the NG Green's function solves (\ref{eq:greeneq}) we can decompose it as
\begin{equation}
G\left(t,\vec{x};t',\vec{x}'\right)=\int\frac{d\omega}{2\pi}e^{-i\omega(t-t')}\sum_{n}\int\frac{d^{2}k}{4\pi^{2}}\frac{ie^{i\vec{k}\left(\vec{x}_{\perp}-\vec{x}'_{\perp}\right)}}{\omega-\omega_{\vec{k},n}+i \mathrm{sgn}(\omega)\epsilon}\chi_{\vec{k},n}(z)\chi_{\vec{k},n}^{+}(z'),
\end{equation}
where $\chi_{\vec{k},n}(z)$ solves the Dirac equation (\ref{eq:eigenvalue})
and form an orthonormal basis
\begin{equation}
\int_{0}^{z_{w}}dz\chi_{\vec{k},n}^{+}(z)\chi_{\vec{k},n'}(z)=\delta_{nn'},
\end{equation}
\begin{equation}
\sum_{n}\chi_{\vec{k},n}(z)\chi_{\vec{k},n}^{+}(z')=\delta(z-z').
\end{equation}

We can immediately perform the $\omega$ integral to get (supposing that $t>t'$)
\begin{equation}
G\left(t,\vec{x};t',\vec{x}'\right)=\sum_{n}\int\frac{d^{2}k}{4\pi^{2}}e^{-i\omega_{\vec{k},n}(t-t')}e^{i\vec{k}\left(\vec{x}_{\perp}-\vec{x}'_{\perp}\right)}\chi_{\vec{k},n}(z)\chi_{\vec{k},n}^{+}(z')\Theta\left(\omega_{\vec{k},n}\right),
\end{equation}
\begin{equation}
G\left(t',\vec{x}';t,\vec{x}\right)=-\sum_{n}\int\frac{d^{2}k}{4\pi^{2}}e^{-i\omega_{\vec{k},n}(t'-t)}e^{i\vec{k}\left(\vec{x}'_{\perp}-\vec{x}_{\perp}\right)}\chi_{\vec{k},n}(z)\chi_{\vec{k},n}^{+}(z')\Theta\left(-\omega_{\vec{k},n}\right).
\end{equation}

Substituting this into (\ref{eq:SGreen}) we obtain (\ref{eq:ASource}) and (\ref{eq:SSourcenonpert}).

\section{Perturbative solution}
\label{sec:pert-solut}

\subsection{Perturbative fermion spectrum, AdS-gap equation}
\label{sec:pert-ferm-spectr}

We will solve the fermionic equation of motion (\ref{eq:eigenvalue}) perturbatively in the scalar interaction and
determine the gap equation.
It is convenient for this to write the eigenvalue problem in terms of $(\alpha_1, \alpha_2, \alpha_3, \alpha_4)=(\chi_1, i \chi_2, \chi_3, i \chi_4)$.
The redefined Hamiltonian is real (but we will still denote it with $H$).

Our Hamiltonian can be split as $H=H_{0}+V$, where $H_{0}=H(\eta_{5}=0)$. The perturbation
is coming from the Majorana coupling
\begin{equation}
V=2\eta_{5}\frac{\phi}{z}\left(\begin{array}{cc}
0 & -\epsilon\\
\epsilon &0
\end{array}\right),
\end{equation}
where $\epsilon$ is the 2x2 matrix $\eps=i\sigma_2$

The solution of the unperturbed problem for a given momentum  takes the form
\begin{equation}
\alpha_{k,+,n}^{(0)}=\left(\begin{array}{c}
\xi_{k,n}\\
0
\end{array}\right)\,\,\omega_{1,k,n}^{(0)}=\omega_{k,n}^{(0)}>0,
\end{equation}
\begin{equation}
\alpha_{k,-,n}^{(0)}=\left(\begin{array}{c}
0\\
\epsilon\xi_{k,n}
\end{array}\right)\,\,\omega_{2,k,n}^{(0)}=-\omega_{k,n}^{(0)}.
\end{equation}
We will focus on $n=1$ and will omit this index.

When doing the perturbation theory we should be careful because
near the Fermi-surface different bands are crossing each other.
Therefore we start with two modes with unperturbed energy
$\omega_{k}^{(0)}$ and $-\omega_{k}^{(0)}$ and approximate the
solution as $\alpha_{k}=a\alpha_{k,+}^{(0)}+b\alpha_{k,-}^{(0)}$.
Near the Fermi-surface this is a good approximation.

The perturbed
energy and wave-functions can be determined by the off-diagonal
matrix element of $V$ (the diagonal elements are zero).
\begin{equation}
V_{k}=\int_{0}^{z_{w}}dz\alpha_{k,+}^{(0)+}V\alpha_{k,-}^{(0)}=2\eta_{5}\int_{0}^{z_{w}}dz \lvert \xi_{k}\rvert^2 \frac{\phi}{z}.
\end{equation}
The new energy levels are
\begin{equation}
\omega_{\pm}=\pm\sqrt{\left(\omega_{k}^{(0)}\right)^{2}+V_{k}^{2}},
\end{equation}
so the size of the gap is $V_{k_{F}}$. The normalized wave-functions
are
\begin{equation}
\alpha_{k,+}=\left(\begin{array}{c}
\xi_{k}\cos\frac{1}{2}\beta_{k}\\
\epsilon\xi_{k}\sin\frac{1}{2}\beta_{k}
\end{array}\right),\,\,\,\,\,\,\alpha_{k,-}=\left(\begin{array}{c}
-\xi_{k}\sin\frac{1}{2}\beta_{k}\\
\epsilon\xi_{k}\cos\frac{1}{2}\beta_{k}
\end{array}\right),
\end{equation}
where
\begin{equation}
\tan\beta_{k}=\frac{V_{k}}{\omega_{k}^{(0)}}.
\end{equation}

Using this perturbative result we can express the scalar source
with the unperturbed fermion wave functions:
\begin{equation}
\langle\bar{\psi^{C}}\Gamma^{5}\psi\rangle=-\frac{i}{4\pi}\int_{-\Lambda
\left(\omega_D\right)}^{\Lambda
\left(\omega_D\right)}dk|k|\frac{V_{k}}{\sqrt{\left(\omega_{k}^{(0)}\right)^{2}+V_{k}^{2}}}\lvert
\xi_{k}(z)\rvert^2.
\end{equation}
Here $\Lambda(\omega_D)$ is a momentum cut-off corresponding to the
energy scale $\omega_D$.
In our numerics we sample discrete number of momenta and
sum over it. In order to capture the contribution around $k_{F}$
accurately we can use the following discretization
\begin{equation}
\int_{-\Lambda \left(\omega_D\right)}^{\Lambda \left(\omega_D\right)}dk|k|\frac{V_{k}}{\sqrt{\left(\omega_{k}^{(0)}\right)^{2}+V_{k}^{2}}}\lvert \xi_{k}\rvert^2\approx\sum_{k_{i}}V_{k_{i}}k_{i}\frac{1}{|\omega'(k_{i})|}\int_{\omega(k_{i})}^{\omega(k_{i+1})}\frac{d\omega}{\sqrt{\omega^{2}+V_{k}^{2}}}
\end{equation}
\[
=\sum_{i}\lvert \xi_{k_i}\rvert^2\frac{V_{k_{i}}k_{i}}{\omega'(k_{i})}\log\left(\frac{\omega_{i+1}^{(0)}+\sqrt{\left(\omega_{i+1}^{(0)}\right)^{2}+V_{k_{i}}}}{\omega_{i}^{(0)}+\sqrt{\left(\omega_{i}^{(0)}\right)^{2}+V_{k_{i}}}}\right)
\]

\subsection{Simplified Gap-equation}
\label{sec:simpl-gap-equat}

The dominant contribution for the scalar charge comes from a
region near the Fermi surface where the (unperturbed) spectrum is
linear. Since the perturbation matrix element $V_{k}$ is a slowly
varying function of $k$ we can approximate its value with
$V_{k_{F,1}}=V_{1}$ and $V_{k_{F,2}}=V_{2}$.

We have two Fermi surfaces. Hence the gap-equation is (recall that our
scalar is an auxiliary field with no dynamics here)
\begin{equation}
\phi(z)=\frac{z^{3}}{4\eta_5}\left[\gamma_1V_1\log\left(\frac{\omega_{D}+\sqrt{\omega_{D}^{2}+V_{1}^2}}{V_{1}}\right)\rho_1(z)+\gamma_2 V_2 \log\left(\frac{\omega_{D}+\sqrt{\omega_{D}^{2}+V_{2}^2}}{V_{2}}\right)\rho_2(z)\right],
\end{equation}
where $\rho_1(z)=\lvert \xi_{k_{F,1}}\rvert^2$,
$\rho_2(z)=\lvert \xi_{k_{F,2}}\rvert^2$ and $\gamma_{1,2}=\frac{\eta_{5}^{2}}{m_{\phi}^{2}\pi}\frac{|k_{F1,2}|}{|\omega'(k_{F1,2})|}$. We make the following
ansatz
\begin{equation}
\phi=\left(C_1\rho_1(z)+C_2\rho_2(z)\right)z^{3}.
\end{equation}
In this case the perturbation matrix element is
\begin{equation}
V_{1}=2\eta_{5}\left(C_1I_{11}+C_2I_{12}\right)
~,~~
V_{2}=2\eta_{5}\left(C_2I_{22}+C_1I_{12}\right),
\end{equation}
where
\begin{equation}
I_{11}=\int_{0}^{z_{w}}z^{2}\rho_{1}^{2}dz,\,\,\,\,\,\, I_{22}=\int_{0}^{z_{w}}z^{2}\rho_{2}^{2}dz,\,\,\,\,\,\, I_{12}=\int_{0}^{z_{w}}z^{2}\rho_2\rho_1dz. \label{eq:I}
\end{equation}
In the limit of $\omega_{D}\gg \eta_{5}$ our gap-equations take the
following form
\begin{align}
aV_{1}+bV_{2}&=2\eta_{5}\gamma_{1}V_{1}\log\left(\frac{\omega_{D}}{\eta_{5}V_{1}}\right)\nonumber\\
bV_{1}+cV_{2}&=2\eta_{5}\gamma_{2}V_{2}\log\left(\frac{\omega_{D}}{\eta_{5}V_{2}}\right),
\end{align}
with
\begin{equation}
a=\frac{I_{22}}{I_{22}I_{11}-I_{12}^{2}} \label{eq:a}
~,~~
b=\frac{I_{12}}{I_{12}^{2}-I_{22}I_{11}}
~,~~
c=\frac{I_{11}}{I_{22}I_{11}-I_{12}^{2}}.
\end{equation}

For the ratio $x=V_{1}/V_{2}$ we obtain
\begin{equation}
x^{2}+\left(\frac{I_{22}}{I_{12}}\frac{\gamma_{2}}{\gamma_{1}}-\frac{I_{11}}{I_{12}}\right)x-\frac{\gamma_{2}}{\gamma_{1}}=\frac{\gamma_{2}}{b}x\log x.
\end{equation}
We can now solve our equations easily to obtain
\begin{equation}
C_1=(ax+b)\frac{\omega_{D}}{\eta_{5}}\exp\left(-\frac{bx+c}{\gamma_{2}}\right)
~,~~
C_2=\left(bx+c\right)\frac{\omega_{D}}{\eta_{5}}\exp\left(-\frac{bx+c}{\gamma_{2}}\right).
\end{equation}

\section{Numerical methods}

\subsection{General strategy}
To solve the equations \eqref{MainSystem1}
numerically, we resort to an iterative Hartree resummation:
\begin{itemize}
 \item At a constant $A_0=\mu$ and zero scalar field, we find the unperturbed spectrum of fermions.
 As a result we get a set of fermionic
 wavefunctions for a discrete array of energies and momenta $(k_i,\omega_{n,i})$.
 \item With these wavefunctions we construct the source terms on the
   right hand side of the first two equations in \eqref{MainSystem1} and solve for $A_0(z)$ and $\phi(z)$. Both UV cut offs in both $k$ and $\omega$
   should be imposed to render the sums in the source terms finite.
 \item Substitute the new $A_0(z)$ and $\phi(z)$ into the Dirac equation and find the new spectrum.
 \item Repeat steps 2 -- 4 till full convergence.
 \end{itemize}
Once the system converges sufficiently, we can extract the information
of the dual theory by a fit to the near boundary behavior of the
resulting wavefunctions.

We have optimized our numerics in several ways:
The most time-consuming part of the algorithm is the repeated
calculation of the Dirac fermion spectrum. A significant improvement
is obtained using the perturbative prescription described in a
previous section. We exclude the $\phi(z)$ field from the Dirac
equation, and instead of four coupled ODE we
get for fermions two identical decoupled systems of a second
order. Then we construct the corrected wavefunctions. In addition, we do not need to take equally dense sampling in $k$, because
most of the fermionic spectral weight is concentrated around $k_F$ (remember that we have two slighly different Fermi momenta in the theory),
and we may take sparser $k$-sampling away from these points without loss in accuracy.

Empirically we found that different numerical schemes to fermionic and
bosonic subsystems was the most efficient. For the fermionic spectrum
we use the shooting method: we impose boundary conditions
dependent on a free parameter at {the boundary cut off} $z=\epsilon$, and scan over this parameter to make the resulting solution satisfy physical boundary conditions at the hard wall.

However, the shooting method in the gauge field and scalar sector often leads the system to converge to some higher harmonics instead of the groundstate.
The Newton method is much more stable in that case: we impose both AdS-infinity and hardwall boundary conditions at the same time, approximate differential
equations by finite differences, and solve the resulting system of
linear algebraic equation with a relaxation algorithm. For our
purposes a grid of $N_p=3000$ points in $z$-direction (for
$z_{w}=1$) was chosen, in which case the relaxation algorithm converges after $5-6$ iterations.

Once the bulk wave functions are obtained, it is still not a trivial
question how to extract the leading boundary behavior from this
data. This is what contains the information of the dual field
theory. The analytical puzzles related to this problem were discussed in section \ref{}. Here we focus on corresponding numerical issues.

We are interested in coefficients ${\cal H}_1,\, {\cal P}_1,\, {\cal P}_3$ defined in \eqref{eq:AlgebraicResonance}.
The function $\phi(z)$ is known in a form of discrete list of values $\left\{z_i,\,\phi(z_i) \right\}$ of the length
$N_p=3000$, therefore our accuracy is limited and naive use of the standard fitting schemes of {\it Mathematica} leads to
large errors.

Instead we first determine the expansion coefficients of the fermionic bilinear
sourcing the scalar field \be
-i\eta_{5}z^{3}\langle\overline{\psi^{c}}\Gamma^{5}\psi\rangle={\cal
S}_1 z^5+{\cal S}_3 z^7+... \ee
These can be easily found, as
contra to the scalar field profile the fermionic bulk wave functions
are derived with a great accuracy due to the use of the shooting
method.

Then we use the algebraic relations \eqref{eq:AlgebraicResonance} to obtain the ``particular'' coefficients on the base of ${\cal S}_i$.

Knowing ${\cal P}_1$ and ${\cal P}_3$ we can subtract these from the scalar wave function and run the Newton relaxation scheme one more time for
\be \tilde{\phi}(z)=\phi(z)-{\cal P}_1z^5-{\cal P}_3z^7.\ee

We now need to fit only for the single coefficient ${\cal H}_1$. This
can be easily done even for moderate number of discretized points $N_p$.

\end{document}